\documentclass[a4paper,floatfix,aps,pra,twocolumn,showpacs,amsmath,preprintnumbers,nofootinbib,secnumarabic,superscriptaddress]{revtex4}

\usepackage{epsfig}
\usepackage{amssymb}
\usepackage{verbatim}
 
\usepackage{graphicx}
\usepackage{color}

\def\bfl{\begin{flushleft}}
\def\efl{\end{flushleft}}
\def\bfr{\begin{flushright}}
\def\efr{\end{flushright}}
\def\bc{\begin{center}}
\def\ec{\end{center}}
\def\be{\begin{equation}}
\def\ee{\end{equation}}
\def\ba{\begin{eqnarray}}
\def\ea{\end{eqnarray}}
\def\baa#1{\begin{array}{#1}}
\def\eaa{\end{array}}
\def\bw{\begin{widetext}}
\def\ew{\end{widetext}}
\def\nn{\nonumber }
\def\lb#1{\label{#1}}
\def\bit{\begin{itemize}}
\def\eit{\end{itemize}}
\def\bco{\begin{comment}} \def\eco{\end{comment}}

\def\drm{d}

\def\schrod{Schr\"odinger  }

\def\Sign#1{\, \text{sign}\left(#1\right) }

\begin{document}

\preprint{\small J. Mod. Optics 61 (2014) 1298-1308 [arXiv:1405.7165]}

\title{
Comparison and unification of non-Hermitian and Lindblad approaches with applications to open quantum optical systems 
}

\author{Konstantin G. Zloshchastiev}
%\email{k.g.zloschastiev@gmail.com}

\affiliation{
School of Chemistry and Physics, University of KwaZulu-Natal, Pietermaritzburg Campus,
Private Bag X01, Scottsville, Pietermaritzburg 3209, South Africa}

\author{Alessandro Sergi}
%\email{sergi@ukzn.ac.za}

\affiliation{
School of Chemistry and Physics, University of KwaZulu-Natal, Pietermaritzburg Campus,
Private Bag X01, Scottsville, Pietermaritzburg 3209, South Africa}

\affiliation{National Institute for Theoretical Physics (NITheP), KwaZulu-Natal Node, South Africa}

\begin{abstract}
We compare two approaches to open quantum systems,
namely, the non-Hermitian dynamics and the Lindblad master equation.
In order to deal with more general dissipative phenomena,
we propose the unified master equation
that combines the characteristics of both of these approaches.
This allows us to 
assess the differences between them
as well as to clarify which observed features
come from the Lindblad or the non-Hermitian part,
when it comes to experiment.
Using a generic two-mode single-atom laser system as
a practical example, we analytically solve the dynamics of
the normalized density matrix operator.
We study the two-level model in a number of cases (depending
on parameters and types of dynamics),
compute different observables
and study their physical properties. 
It turns out that one is able not
only to describe the different types of damping in dissipative quantum optical systems
but also to mimic the
undamped anharmonic oscillatory phenomena which happen in quantum systems with more than two levels
(while staying within the framework of the analytically simple two-mode approximation).
\end{abstract}

\date{\footnotesize Received: 2 April 2014 [JMO], 28 May 2014 [arXiv]}

%\today

\maketitle

\section{Introduction}

There are various theoretical methods that can be used in order to study open quantum systems.
One can start by considering the totality of the degrees of freedom and afterwards focus, either
analytically \cite{attal} or numerically \cite{ilya}, on a subset of relevant coordinates.
However, the approach that is currently more popular, immediately integrates over
the degrees of freedom of the environment and, for the Markovian systems,
establishes the Lindblad master equation \cite{gks76,lin76}.
Originally introduced in spin physics and quantum optics \cite{bau66,haa73}, master equations 
have been applied to many dissipative quantum phenomena \cite{carbook,gzbook,bpbook}.
 
There is yet another approach to open quantum systems that is based on a
non-Hermitian extension of quantum mechanics (NHQM) \cite{fesh58,wong67,faisa,datto,heg93,ang95,rotter,gsz08,grsc11,sz12,bg12,ghk10,thila}.
In this approach the Hamiltonian is assumed 
to acquire an anti-Hermitian part which can be associated with dissipative 
effects.
Both the Lindblad and non-Hermitian approaches are based on 
certain simplifying assumptions
and have their own range of applicability.
Moreover, one can find different physical motivations and theoretical advantages
in their respective use:
the Lindblad master equation is linear and permits the simple calculation
of averages whereas non-Hermitian dynamics possesses a generalized canonical structure 
that leads to a well-defined classical limit \cite{ghk10}.
%pippo

In the present work, we compare the evolution arising from the
Lindblad approach to the one from the non-Hermitian approach. 
Moreover, we propose a ``hybrid'' formalism that combines features of both of them,
and, as such, it is expected to have a wider range of applicability.
In order to illustrate the theory, we  consider quantum two-level system (TLS)
and study its  time evolution 
while mimicking the coupling to a dissipative environment through
non-Hermitian dynamics, Lindblad master equation, or their combination.
The dynamics of the density matrix is solved analytically
in a number of relevant cases, the behavior of primary observables (averages)
is studied accordingly.

This paper is structured as follows.
In section \ref{s-two} we give a brief outline of the Lindblad and non-Hermitian approaches.
In section \ref{s-uni} we combine them into a unified approach, named
``hybrid'' throughout the paper,
and adopt a two-level system as a practical example.
In sections \ref{s-gsol}, \ref{s-lc} and \ref{s-apsol} we consider
different limits of the hybrid dynamics of the two-level model, derive
their analytical solutions and calculate the relevant observable properties. 
Some facts about two-level systems in quantum optics and relevant notations are reminded in the Appendix \ref{s-a-tls}.
The definitions of the Fourier transforms used in the paper are given in the Appendix \ref{s-a-ft}.
Conclusions are drawn in section \ref{s-con}.

%\newpage

\section{Open quantum system dynamics}\lb{s-two}

In this section we give a brief description
of two popular density-operator based approaches which
are used in a theory of open quantum systems.
In both approaches, one distinguishes the coordinates
of a subsystem from those of the environment so that the total
Hamiltonian can be written as the sum $H_{\rm T} = H_{\rm S}+H_{\rm B}$ 
of the Hamiltonians of the relevant subsystem, $H_{\rm S}$, and
of the environment (or bath), $H_{\rm B}$.
Accordingly, the total Hilbert space
becomes the product of two composing Hilbert spaces,
\be
{\cal H}_\text{T} = {\cal H}_{\rm S} \otimes {\cal H}_{\rm B}\; ,
\ee
where ${\cal H}_{\rm S}$ and ${\cal H}_{\rm B}$
are the Hilbert spaces of the relevant subsystem and bath, respectively.
The density operator of the relevant subsystem is obtained
by tracing out the degrees of freedom of the environment
from the total density matrix:
\be\lb{eq:reddo}
\hat\rho_{\rm S} \equiv \hat\rho 
= {\rm tr}_B \, \hat{\rho}_\text{T} 
,
\ee
where ${\rm tr}_{\rm B}$ denotes the partial trace over the degrees of freedom of the
environment $B$.
Despite the common basis, 
the Lindblad and non-Hermitian approaches
describe different effects of the environment onto the subsystem 
and produce different properties of the latter.

\subsection{Lindblad master equation}\lb{s-meq}

This approach, while maintaining a Hamiltonian contribution 
to the dynamics of the subsystem, 
modifies the evolution equation of the subsystem density operator
by adding dissipative terms (\ref{eq:reddo}).
Upon using the Markov approximation and some auxiliary simplifications,
one can show that the Lindblad master equation takes the form:
\begin{equation}
\frac{\drm}{\drm t}\hat{\rho}(t)
=-\frac{i}{\hbar}\left[\hat{H}_+, \hat{\rho}(t)\right]
+
\hat{\cal D} ({\rho}(t),  A_k )
,
\label{eq:master1}
\end{equation}
where
$\hat H_+ = \hat H_+^\dagger$ is a Hamiltonian
that commutes with that of the subsystem $\hat{H}_{\rm S}$.
The  dissipator $\hat{\cal D} ({\rho},  A_k )$  in Eq.~(\ref{eq:master1}) is 
a linear operator in $\hat{\rho}(t)$ and quadratic in the Lindblad operators $\hat A_k$,
$k = 1,..., N^2-1$ and $N = \text{dim} ({\cal H}_S)$.
The dissipator must be traceless for the trace of the density operator to be
conserved during evolution.
The Lindblad operators $A_k$ describe directly the dissipative effects of the
environment.
The most general quantum dynamical semigroup form of the dissipator~\cite{bpbook}
can be written as
\be
\hat{\cal D} ({\rho},  A_k ) =
\sum\limits_{k=1}^{N^2-1} \gamma_k 
\left(
\hat A_k \hat{\rho} \hat A_k^\dagger
-
\frac{1}{2}
\hat A_k^\dagger \hat A_k \hat{\rho}
-
\frac{1}{2}
\hat{\rho} \hat A_k^\dagger \hat A_k
\right)
,
\ee
where $\gamma$'s are non-negative quantities which can be expressed in terms of certain correlation
functions of the environment and play the role of relaxation rates for different decay modes
of the open subsystem. 

Simple examples of the application of the Lindblad master equation are found 
when studying models of two-level
atoms interacting with the electromagnetic field in presence of a thermal 
reservoir of radiation modes~\cite{carbook,gzbook,bpbook}.

Using the notation of Appendix~\ref{s-a-tls}, we can write the master equation in
the interaction picture as
\be
\frac{\drm}{\drm t}\hat{\rho}_{(I)} (t)
=
\frac{i \Omega}{2}
\left[\hat{\sigma}_+ + \hat{\sigma}_- , \hat{\rho}_{(I)} (t) \right]
+
\hat{\cal D} ({\rho}_{(I)} (t))
,
\label{eq:blocheq}
\ee
where the dissipator is given by
\bw
\ba
&&
\hat{\cal D} ({\rho})
= 
\gamma_0 (N+1)
\left[
\hat{\sigma}_- \hat{\rho} \hat{\sigma}_+ -
\frac{1}{2}
\{\hat{\sigma}_+\hat{\sigma}_-, \hat{\rho}\}
\right]
+
\gamma_0 N
\left[
\hat{\sigma}_+ \hat{\rho} \hat{\sigma}_- -
\frac{1}{2}
\{\hat{\sigma}_-\hat{\sigma}_+, \hat{\rho}\}
\right]
,
\lb{eq:disstls}
\ea
\ew
where $\gamma_0$  is the spontaneous emission rate and
$N = N(\omega_0)$ denotes the Planck distribution at the transition frequency \cite{bpbook}.

\subsection{Non-Hermitian approach}\lb{s-nhd}

Non-Hermitian dynamics has found various applications in the study of open quantum system \cite{fesh58,wong67,faisa,datto,heg93,ang95,rotter,gsz08,grsc11,sz12,bg12}.
Recently, it has been shown \cite{sz12} that this approach is capable of 
describing the evolution of pure states
into mixed ones - since the purity is not necessarily conserved 
if the dimension of the corresponding Hilbert space is larger than two.
In turn, such a feature can be used in the quantitative modeling of the observer-related 
phenomena in quantum mechanics - such as the problem of measurement 
and phenomenon of decoherence.
In the simplest formulation of this approach, it is assumed that the dissipative effects of the
environment are somehow encoded in anti-Hermitian terms of the subsystem
Hamiltonian which appear after averaging (``integrating out'') the degrees of freedom
of environment.
Hence, in this approach one deals exclusively with the degrees of freedom of the 
subsystem, which is in turn described by a non-Hermitian Hamiltonian.

The non-Hermitian Hamiltonian operator can be always 
partitioned into Hermitian and anti-Hermitian parts
\be
\hat{H} = \hat{H}_+ +\hat{H}_- , \lb{e-hplusmin}
\ee
where we denoted $\hat{H}_{\pm}= \pm \hat{H}_{\pm}^{\dag}
= \tfrac{1}{2} (\hat H \pm \hat H^\dag)$.
For further it is convenient to introduce also the self-adjoint operator
$\hat\Gamma \equiv  i \hat H_-$ which
will be referred as the \textit{decay rate operator} throughout the paper.
Starting from the Schr\"odinger equation,
it is easy to show that the evolution equation for the density operator acquires 
an anticommutator term:
\begin{equation}
\frac{\drm}{\drm t}\hat{\rho}(t)
=-\frac{i}{\hbar}\left[\hat{H}_+,\hat{\rho}(t)\right]
-\frac{i}{\hbar}\left\{\hat{H}_-,\hat{\rho}(t)\right\}
.
\label{eq:dtrho}
\end{equation}
Equation (\ref{eq:dtrho}) can also be written in matrix form~\cite{sergi-mat},
directly implementing, within a quantum framework, the original geometric ideas
of Grmela \cite{grmela} about dissipation.

While equations of the form (\ref{eq:dtrho}) find some applications \cite{heg93,grsc11},
they also possess certain features (which also manifest themselves when working
with the state vectors) that narrow their applicability range.
For instance, 
the trace of the density operator determined by such equations
is not preserved in general:
\be\lb{e-trrhorate}
\frac{\drm}{\drm t}
{\rm tr} \left(\hat{\rho} (t) \right)
=
\frac{2}{i \hbar}
{\rm tr} \left(\hat{\rho} (t) \hat H_-  \right)
.
\ee
This renders the usual probabilistic interpretation of quantum mechanics
more problematic to achieve.
Another issue arises if one studies the invariance of the evolution equation under 
the Hamiltonian ``gauge'' shift
\be\lb{eq:gaugeiunit}
\hat H
\rightarrow \hat H +  c_0 \hat I,
\ee
where $c_0$ is a c-number and $\hat I$ is an identity operator. 
As it happens in conventional quantum mechanics,
one would like that
such a shift should affect neither the observable averages nor the evolution equation.
However, according to (\ref{eq:dtrho}), this invariance gets broken if $c_0$ has an imaginary part.

In view of these circumstances, in our previous work \cite{sz12} we proposed to
consider the normalized density operator,
\be\lb{e-dmatnorm}
\hat\rho' = 
\hat\rho / {\rm tr} \left(\hat{\rho} \right)
,
\ee
as a primary physical object of theory.
Following this idea, 
the quantum average of an observable $\hat{O}=\hat{O}(0)$ 
is defined in terms of the normalized density operator in Eq.~(\ref{e-dmatnorm}) as
\begin{equation}
\langle O\rangle_\text{obs}
\equiv
{\rm tr}\left(\hat{\rho}' \hat{O}(0)\right)
=
{\rm tr}\left(\hat{\rho}\hat{O}(0)\right)/ {\rm tr}\left(\hat{\rho}(t)\right)
.
\label{eq:spicture}
\end{equation}
This idea was also adopted in \cite{bg12} where the 
evolution equation, which can be derived for the normalized density operator
in our approach, 
was favored over the equations for the non-normalized operator 
and state vectors which were used previously (cf. \cite{grsc11}, for instance).
It turns out that the normalized density operator approach automatically solves 
the above-mentioned issues of norm-conservation and gauge invariance:
using the evolution equation
which follows from (\ref{eq:dtrho}) 
and (\ref{e-dmatnorm}),
\bw
\ba
&&
%i \hbar
\frac{\drm}{\drm t}\hat{\rho}' (t)
=
-\frac{i}{\hbar}
\left[\hat{H}_+,\hat{\rho}' (t)\right]
-\frac{i}{\hbar}
\left\{\hat{H}_-,\hat{\rho}' (t)\right\}
%+ 2 i\,
%\nn\\&&\qquad \qquad \qquad
%+
%i \hbar
%\hat{\cal D} ({\rho}' (t),  A_k ) 
+ \frac{2 i}{\hbar}
%-2 \,
{\rm tr} \left(\hat{\rho}' (t) \hat H_-  \right)
\hat{\rho}' (t)
,
\label{eq:evnorm}
\ea
\ew
one can easily check that the normalization property ${\rm tr} (\hat{\rho}') = 1$ is conserved
and that
the ``gauge'' invariance under the transformation (\ref{eq:gaugeiunit}) 
is achieved for arbitrary $c_0$.

To conclude, the main advantage of the normalized density-operator approach in NHQM is
that it handles in a unified way not only the pure states
but also the mixed ones.
Moreover, the emerging nonlinearity of the evolution equation (\ref{eq:evnorm})
provides yet another example of a profound interplay between
the physics
of open quantum systems and nonlinear quantum mechanics:
the environment is able to induce
effective nonlinearities in quantum evolution equations
\cite{gisin,gisin2,gisin3,ks87,various1,various2,various3,various4,various5,various6,various7,various8,various9,various10,az11,zlo2012}.

\section{``Hybrid'' formalism}\lb{s-uni}

In this section we unify the Lindblad master equation with the non-Hermitian
equation for the density matrix.
Such a hybrid equation is postulated 
replacing the usual Hamiltonian contribution to the evolution of the
non-normalized density matrix of the
quantum subsystem in the Lindblad master equation with a more general
non-Hermitian one, taken from the NH equation (\ref{eq:dtrho}).
In such a way one obtains the following equation
\begin{equation}
\frac{\drm}{\drm t}\hat{\rho}(t)
=-\frac{i}{\hbar}\left[\hat{H}_+,\hat{\rho}(t)\right]
-\frac{i}{\hbar}\left\{\hat{H}_-,\hat{\rho}(t)\right\}
+
\hat{\cal D} ({\rho}(t),  A_k )
,
\label{eq:hybevnnorm}
\end{equation}
where 
\begin{equation}
\hat{H} = \hat{H}_+ +\hat{H}_- = \hat{H}_+ - i \hat\Gamma . \lb{eq:hplusmin2}
\end{equation}
Upon substituting the normalized density operator (\ref{e-dmatnorm})
into equation (\ref{eq:hybevnnorm}), one obtains a
a nonlinear evolution equation,
\bw
\ba
&&
%i \hbar
\frac{\drm}{\drm t}\hat{\rho}' (t)
=
-\frac{i}{\hbar}
\left[\hat{H}_+,\hat{\rho}' (t)\right]
-\frac{i}{\hbar}
\left\{\hat{H}_-,\hat{\rho}' (t)\right\}
%+ 2 i\,
%\nn\\&&\qquad \quad 
+
%i \hbar
\hat{\cal D} ({\rho}' (t),  A_k )
+ \frac{2 i}{\hbar}
%-2 \,
{\rm tr} \left(\hat{\rho}' (t) \hat H_-  \right)
\hat{\rho}' (t)
.
\label{eq:hybevnorm}
\ea
\ew
Below it will be shown that this nonlinearity makes the models
based on the hybrid equations
substantially more interesting than those obtained from
the Lindblad or non-Hermitian equations alone.

In what follows, we study a two-level optical quantum system which is both
an instructive example and a fruitful physical application.
Using the notation and the definition of the system given in Appendix \ref{s-a-tls},
we assume that the evolution
is governed by equations (\ref{eq:hybevnnorm}), (\ref{eq:hplusmin2}) and (\ref{eq:disstls}).
The model we study is defined by the following Hermitian Hamiltonian
\begin{eqnarray}
\hat H_+ &=& \hat H_0 + \hat H_L ,
\end{eqnarray}
where
\begin{eqnarray}
\hat H_0 &=& \frac{1}{2} \hbar \omega_0 \hat{\sigma}_3 ,
\\
\hat H_L &= & \frac{1}{2} \hbar \Omega 
\left( \text{e}^{-i \omega_0 t} \hat{\sigma}_+
+ \text{e}^{i \omega_0 t} \hat{\sigma}_-
\right) 
.
\end{eqnarray}
The unperturbed Hamiltonian $\hat H_0 $ can represent the two energy levels of a free dipole.
In such a case, the perturbation $\hat H_L$ would describe the interaction between
the dipole and a single-mode electromagnetic wave.
%The physical meaning of this Hamiltonian
More details and corresponding notations are provided in
the Appendix \ref{s-a-tls}.

The anti-Hermitian Hamiltonian, which must be added to $\hat H_+$ to give
the total Hamiltonian  $\hat H$ of the model, is
\be
\hat H_- =  \hat H_\Gamma + \hat H_{D} +  \hat H_{00}
,
\ee
where
\ba
\hat H_\Gamma &=& \frac{1}{2} i \hbar \Gamma \hat{\sigma}_3, \\
\hat H_D &=&
- \frac{1}{2} i \hbar \alpha \left( \text{e}^{-i \omega_0 t} \hat{\sigma}_+
+ \text{e}^{i \omega_0 t} \hat{\sigma}_- \right)
, \\
\hat H_{00} &=& - \frac{1}{2} i \hbar {\cal T} \hat{I}
,
\end{eqnarray}
where $\Gamma$, $\alpha $ and ${\cal T}$ are real-valued free parameters.
The Hamiltonian $\hat H_\Gamma \propto \hat{\sigma}_+ \hat{\sigma}_- $
is motivated by the physics of photodetection
and continuous measurements in presence of radiation modes, cf. Sec. 6.3.1 of \cite{bpbook}.
The term $\hat H_D$ is the anti-Hermitian counterpart of $\hat H_L $, therefore, it is supposed to
describe the dissipative processes accompanying the dipole interaction of the atom and electromagnetic field.
The term $\hat H_{00}$ is a ``gauge'' term - as mentioned in the section \ref{s-nhd},
it does not affect observable values (as defined by
(\ref{e-dmatnorm}) and (\ref{eq:spicture})); however,
it can be used for simplifying or regularizing intermediate expressions.

In terms of the above, the evolution equation of the model
in the interaction picture reads
\ba
&&
\frac{\drm}{\drm t}\hat{\rho}_{(I)} 
=
\frac{i \Omega}{2} 
\left[\hat{\sigma}_+ + \hat{\sigma}_- , \hat{\rho}_{(I)} \right]
-\frac{\alpha}{2} 
\left\{\hat{\sigma}_+ + \hat{\sigma}_- , \hat{\rho}_{(I)} \right\}
\nn\\&&\qquad\quad
+
\gamma_0 (N+1)
\left(
\hat{\sigma}_- \hat{\rho}_{(I)} \hat{\sigma}_+ -
\frac{1}{2}
\{\hat{\sigma}_+\hat{\sigma}_-, \hat{\rho}_{(I)} \}
\right)
\nn\\&&\qquad\quad
+
\gamma_0 N
\left(
\hat{\sigma}_+ \hat{\rho}_{(I)}  \hat{\sigma}_- -
\frac{1}{2}
\{\hat{\sigma}_-\hat{\sigma}_+, \hat{\rho}_{(I)} \}
\right)
\nn\\&&\qquad\quad
+
\frac{\Gamma}{2} 
\left\{\hat{\sigma}_3, \hat{\rho}_{(I)} \right\}
-
{\cal T} \hat{\rho}_{(I)}
.
\label{eq:hybevnnorm2}
\ea
%and equations (\ref{e-dmatnorm}) and (\ref{eq:spicture}) must be kept in mind.
It is convenient to search for solutions of this equation in the form
\ba
&&
\hat\rho_{(I)} (t)=
\frac{1}{2}
\left[
\hat I {\rm tr} (\hat\rho_{(I)} (t))
+
\sum\limits_{i=1}^3 
\hat \sigma_i \langle \sigma_i (t)\rangle_{(I)}
\right]
\nn\\&&\qquad
=
\frac{1}{2}
\left[
\hat I {\rm tr} (\hat\rho_{(I)} (t))
+
\hat\sigma_3 \langle \sigma_3 (t)\rangle_{(I)}
\right]
\nn\\&&\qquad \quad
+ 
\hat\sigma_+ \langle \sigma_- (t)\rangle_{(I)}
+
\hat\sigma_- \langle \sigma_+ (t)\rangle_{(I)}
,
\lb{eq:rhoidecomp}
\ea
where 
\be
\langle \sigma_i (t)\rangle_{(I)} = {\rm tr} (\hat\sigma_i \hat\rho_{(I)} (t)) ,
\ee 
with $i=1,...,3$,
are auxiliary average values that are regarded as unknown functions of time,
together with ${\rm tr} (\hat\rho_{(I)} (t)) = {\rm tr} (\hat\rho (t))$.
The equations for the average values easily follow from equation (\ref{eq:hybevnnorm2})
\ba
&&
\frac{\drm}{\drm t}
\langle \vec\sigma (t)\rangle_{(I)}
=
{\bf G}
\langle \vec\sigma (t)\rangle_{(I)}
+ 
\vec b \,
{\rm tr} (\hat\rho_{(I)} (t)) 
,
\lb{eq:aveint1a}
%\ee\be
\\&&
\frac{\drm}{\drm t}
{\rm tr} (\hat\rho_{(I)} (t))
=
- \alpha \langle \sigma_1 (t)\rangle_{(I)}
+ \Gamma \langle \sigma_3 (t)\rangle_{(I)}
\nn\\&&\qquad\qquad\qquad \ \ \,
- {\cal T} {\rm tr} (\hat\rho_{(I)} (t)) 
,
\lb{eq:aveint1b}
\ea
where we have introduced the matrix
\begin{equation}
{\bf G}
=
\left(
\begin{array}{ccc} 
-\frac{1}{2} \gamma - {\cal T} & 0 & 0 \\
0& -\frac{1}{2} \gamma - {\cal T} & \Omega \\
0& -\Omega & - \gamma - {\cal T} 
\end{array}
\right)
, 
\end{equation}
and  three-dimensional vector
\begin{equation}
\vec b
=
\left(
\begin{array}{c} 
-\alpha \\
0 \\
\Gamma - \gamma_0
\end{array}
\right)
.
\end{equation}
We have also adopted the vector notation $\vec\sigma =(\sigma_1,\sigma_2,\sigma_3)$.
It is very convenient to combine equations~(\ref{eq:aveint1a}) 
and (\ref{eq:aveint1b}) into a single matrix equation 
for the four unknown functions, 
$\langle \sigma_\mu (t)\rangle_{(I)} \equiv 
\left\{ \langle \vec\sigma (t)\rangle_{(I)}, {\rm tr} (\hat\rho_{(I)} (t)) \right\}$
($\mu=1,...,4$):
\be
\frac{\drm}{\drm t}
\langle \sigma_\mu (t)\rangle_{(I)}
=
{\bf M}
\langle \sigma_\mu (t)\rangle_{(I)}
,
\lb{eq:aveint2}
\ee
where 
\bw
\begin{equation}
{\bf M}
=
\left(
\begin{array}{cccc} 
-\frac{1}{2} \gamma - {\cal T} & 0 & 0 &-\alpha \\
0& -\frac{1}{2} \gamma - {\cal T} & \Omega &0 \\
0& -\Omega & - \gamma - {\cal T} & \Gamma - \gamma_0\\
-\alpha& 0& \Gamma & - {\cal T} 
\end{array}
\right)
, 
\;
\hat\sigma_\mu
=
\left(
\begin{array}{c} 
\hat\sigma_1 \\
\hat\sigma_2\\
\hat\sigma_3\\
\hat I
\end{array}
\right)
.
\end{equation}
Equation (\ref{eq:aveint2}) is in a matrix form that is useful in order to
search for solutions.
Nevertheless, it is instructive to re-write it in terms of normalized averages.
Using equations~(\ref{e-dmatnorm}) and (\ref{eq:rhoidecomp}), one can write the normalized density operator in a
decomposed form
\be
\hat\rho_{(I)}' (t)
=
\frac{1}{2}
\left[
\hat I 
+
\sum\limits_{i=1}^3 
\hat \sigma_i \langle \sigma_i' (t)\rangle_{(I)}
\right]
=
\frac{1}{2}
\left[
\hat I 
+
\hat\sigma_3 \langle \sigma_3' (t)\rangle_{(I)}
\right]
+ 
\hat\sigma_+ \langle \sigma_-' (t)\rangle_{(I)}
%\nonumber\\ &+&
+
\hat\sigma_- \langle \sigma_+' (t)\rangle_{(I)}
,
\lb{eq:normdodecomp}
\ee
%\ew
where
\begin{eqnarray}
\langle \sigma_i' (t)\rangle_{(I)} &=& {\rm tr} (\hat{\sigma}_i \hat\rho_{(I)}' (t))
%\nonumber\\ &=& 
= \langle \sigma_i (t)\rangle_{(I)}/ {\rm tr} ( \hat\rho (t))
,
\lb{eq:avernorm}
\end{eqnarray} 
($i=1,...,3$)
are the observable average values in the interaction picture
which satisfy the equation
\ba
&&
\frac{\drm}{\drm t}
\langle \vec\sigma' (t)\rangle_{(I)}
=
{\bf G}_\text{eff} (t)
\langle \vec\sigma' (t)\rangle_{(I)}
+ 
\vec b 
%\, {\rm tr} (\hat\rho_{(I)} (t)) 
,
\lb{eq:aveint3a}
\ea
where
%\bw
\be
{\bf G}_\text{eff} (t)
=
{\bf G}
+
\left(
%\alpha \langle \sigma_1' (t)\rangle_{(I)} - \Gamma \langle \sigma_3' (t)\rangle_{(I)}
{\cal F}_{(I)}(t)
+ {\cal T} 
\right)
\hat I
=
\left(
\begin{array}{ccc} 
{\cal F}_{(I)}(t) -\frac{1}{2} \gamma & 0 & 0 \\
0&{\cal F}_{(I)}(t) -\frac{1}{2} \gamma  & \Omega \\
0& -\Omega &{\cal F}_{(I)}(t) - \gamma 
\end{array}
\right)
,
%\\&& {\cal F}_{(I)}(t) \equiv \alpha \langle \sigma_1' (t)\rangle_{(I)} - \Gamma \langle \sigma_3' (t)\rangle_{(I)} ,
\ee
\ew
and 
${\cal F}_{(I)}(t) \equiv \alpha \langle \sigma_1' (t)\rangle_{(I)} - \Gamma \langle \sigma_3' (t)\rangle_{(I)}$.
While this form of evolution equation is somewhat unsuitable for searching for
analytical solutions
(due to its non-linearity with respect to unknown functions $\langle \sigma_i' (t)\rangle_{(I)}$),
it allows us to demonstrate a feature mentioned in section \ref{s-nhd}:
the contribution from the ``gauge'' term $\hat H_{00}$ disappears when one deals with observable values.

\section{General solution}\lb{s-gsol}

In this section, we search for solutions of Eq.~(\ref{eq:aveint2}) 
in the zero-temperature limit. This is equivalent to setting
$
\gamma = \gamma_0
.
$
Imposing the gauge condition
$
{\cal T} = 0
$
and rescaling the time variable 
$
\tau = \Omega t ,
$
we reduce (\ref{eq:aveint2}) to the form:
\bw
\be
\frac{\drm}{\drm \tau}
\langle \sigma_\mu (\tau) \rangle_{(I)}
=
\widetilde{\bf M}
\,
\langle \sigma_\mu (\tau) \rangle_{(I)}
,
\lb{eq:aveintsol}
\ee
($\mu = 1,...,4$)
where 
\begin{equation}
\widetilde{\bf M}
=
\left(
\begin{array}{cccc} 
-2 \tilde\gamma_0  & 0 & 0 &-\tilde\alpha \\
0& -2 \tilde\gamma_0  & 1 &0 \\
0& -1 & -4 \tilde\gamma_0  & \tilde\Gamma - 4 \tilde\gamma_0\\
-\tilde\alpha& 0& \tilde\Gamma & 0 
\end{array}
\right)
, 
\
\left\langle \hat\sigma_\mu (\tau) \right\rangle
=
\left(
\begin{array}{c} 
\left\langle \hat\sigma_1 (\tau) \right\rangle \\
\left\langle \hat\sigma_2 (\tau) \right\rangle\\
\left\langle \hat\sigma_3 (\tau) \right\rangle\\
{\rm tr} \rho (\tau)
\end{array}
\right)
,
\end{equation}
\ew
In the definition of $\widetilde{\bf M}$, we have introduced the symbols
$\tilde\gamma_0 = \gamma_0/ 4 \Omega$,
$\tilde\alpha = \alpha/ \Omega$
and
$\tilde\Gamma = \Gamma/ \Omega$.
As in the previous sections, one
should keep in mind that the non-normalized
values $\langle \sigma_\mu\rangle$
are auxiliary quantities that are used for computing
the observables, equations~(\ref{eq:normdodecomp}) and (\ref{eq:avernorm}).  

The solution of equation~(\ref{eq:aveintsol}) can be formally written
in a matrix exponential form:
\be\lb{eq:gensolA}
\langle \sigma_\mu (\tau) \rangle_{(I)}
=
%\text{e}
e^{\widetilde{\bf M} \tau}
\langle \sigma_\mu (0) \rangle_{(I)}
,
\ee
where
\be
\langle \sigma_\mu (0) \rangle_{(I)}
=
\langle \sigma_\mu (0) \rangle
=
\langle \sigma_\mu' (0) \rangle_{(I)}
=
\langle \sigma_\mu' (0) \rangle
\lb{eq:ini4values}
\ee
are initial values.
We used the property $\hat\rho (0)= \hat\rho' (0)$,
which is valid
both in the \schrod and in the interaction picture.
Naturally, one also finds that
\be
\langle \sigma_4 (0) \rangle_{(I)}
=
\langle \sigma_4 (0) \rangle
=
\langle \sigma_4' (0) \rangle_{(I)}
%\nonumber\\ &=&
\langle \sigma_4' (0) \rangle =
{\rm tr} \hat\rho (0) = 1
,
\ee
for all physical situations.

Furthermore, if the matrix $\widetilde{\bf M}$ is diagonalizable,
the general solution (\ref{eq:gensolA}) 
can be written in the more convenient form
\ba
&&
\langle \sigma_\mu (\tau) \rangle_{(I)}
=
%|| \text{diag} (\exp{M_{(1)} \tau}) ||
{\bf S}
\langle \sigma_\mu (0) \rangle_{(I)}
,
\lb{eq:gensolB}\\&&
{\bf S}
=
{\bf P}
\left(
\baa{cccc}
e^{M_{(1)} \tau }& 0 & 0 & 0\\
 0 & e^{M_{(2)} \tau } &0 & 0\\
 0 & 0 &e^{M_{(3)} \tau }& 0\\
 0 & 0 & 0 &e^{M_{(4)} \tau } 
\eaa
\right)
{\bf P}^{-1}
, 
\ea
where $M_{(\mu)}$ are eigenvalues of $\widetilde{\bf M}$,
and  ${\bf P}$ is the matrix whose columns are the eigenvectors of $\widetilde{\bf M}$. 
The four-by-four matrix $\widetilde{\bf M}$, with its four eigenvalues,
arises from the hybrid master equation. Such an equation is defined
in terms of 
both the dissipator and the total Hamiltonian
$\hat H_+ + \hat H_-$ (which has just two eigenvalues for the model under study).
Hence, the four eigenvalues of $\widetilde{\bf M}$ carry more physical
information about the studied model system
than the information carried by the eigenvalues of the total Hamiltonian alone. 

\begin{figure}[htbt]
\begin{center}\epsfig{figure=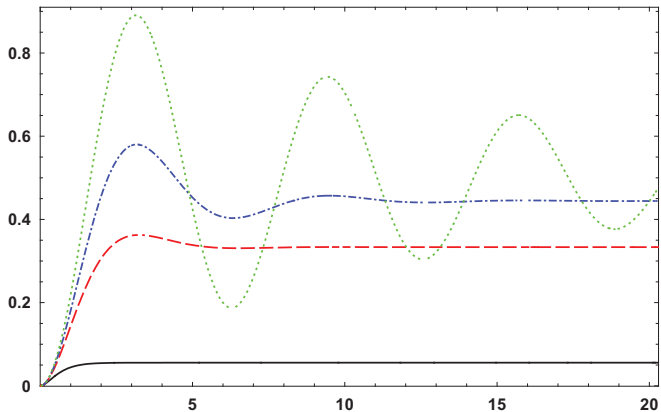,width=1.01\columnwidth}\end{center}
\caption{
The population of the upper level $p_e$ as a function of time 
$\tau = \Omega t$ for the parameters $\tilde\alpha =\tilde\Gamma =0$ and:
$\tilde\gamma_0 = 1$ (solid curve),
$\tilde\gamma_0 = 1/4$ (dashed curve),
$\tilde\gamma_0 = 1/8$ (dash-dotted curve),
and $\tilde\gamma_0 = 1/40$ (dotted curve).}
\label{f-bp-pe}
\end{figure}

\begin{figure}[htbt]
\begin{center}\epsfig{figure=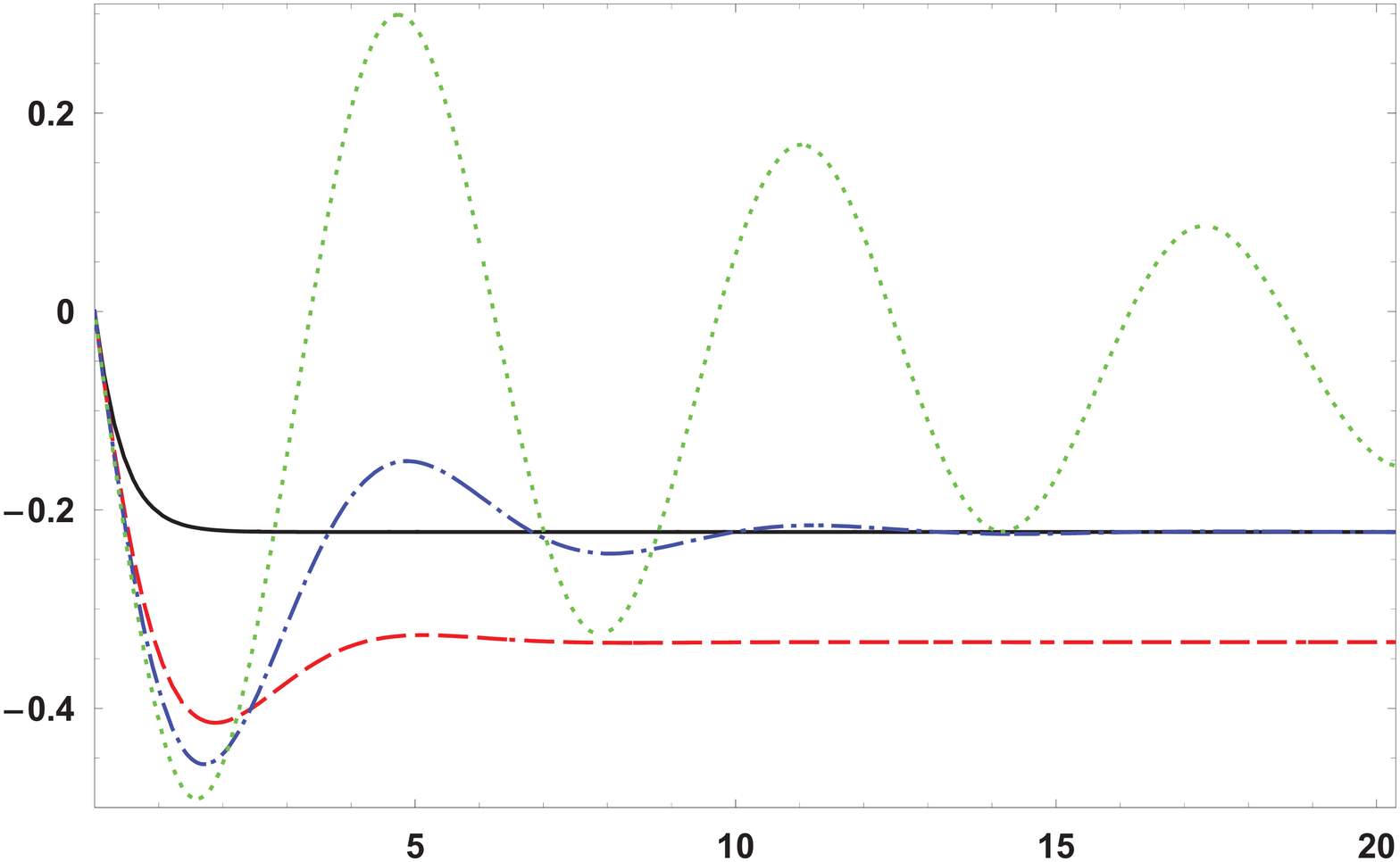,width=1.01\columnwidth}\end{center}
\caption{
The imaginary part of the coherence, $\text{Im} \left(\left\langle {\sigma}_+ \right\rangle_\text{obs} \right)$, 
as a function of time 
$\tau = \Omega t$ for the parameters $\tilde\alpha = \tilde\Gamma =  0$ and:
$\tilde\gamma_0 = 1$ (solid curve),
$\tilde\gamma_0 = 1/4$ (dashed curve),
$\tilde\gamma_0 = 1/8$ (dash-dotted curve),
and $\tilde\gamma_0 = 1/40$ (dotted curve).}
\label{f-bp-sigy}
\end{figure}

\section{Limit cases}\lb{s-lc}

In this section we consider  two special (limit) cases of the general
solution found in section \ref{s-gsol}.
In particular, we treat the case
when dissipative effects are modeled either
only by the Lindblad term (i.e., $\alpha = \Gamma =0$)
and the case when
only the anti-Hermitian term (i.e., $\gamma_0 =0$) is present.
This will allow us to obtain a clearer understanding of the
capabilities of the hybrid formalism.

\begin{figure}[htbt]
\begin{center}\epsfig{figure=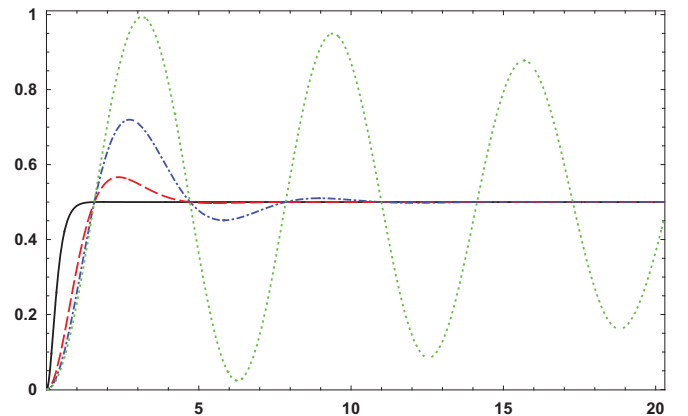,width=1.01\columnwidth}\end{center}
\caption{
The population of the upper level $p_e$ as a function of time 
$\tau = \Omega t$ for the parameters $\tilde\gamma_0 =\tilde\Gamma = 0$ and:
$\tilde\alpha = 4$ (solid curve),
$\tilde\alpha = 1$ (dashed curve),
$\tilde\alpha = 1/2$ (dash-dotted curve),
and $\tilde\alpha = 1/20$ (dotted curve).}
\label{f-g0-pe}
\end{figure}

\subsection{Lindblad-driven dissipation}\lb{s-lc-li}

When $\alpha = \Gamma =0$, the matrix $\widetilde{\bf M}$
has the following four eigenvalues:
\bw
\be
M_1^{(L)} = - 2 \tilde\gamma_0, \
M_2^{(L)} =- 3 \tilde\gamma_0 - i \kappa, \
M_3^{(L)} =- 3 \tilde\gamma_0 + i \kappa,  \
M_4^{(L)}=0,
\ee
where the quantity $\kappa = \sqrt{1- \tilde\gamma_0^2}$ can be imaginary or real-valued.
Correspondingly, the solution from Sec. \ref{s-gsol}
can be written as
\ba
&&
\langle \sigma_\mu (\tau) \rangle_{(I)}
=
{\bf S}_{(L)}
\langle \sigma_\mu (0) \rangle_{(I)}
,
\lb{eq:gensolL}\\&&
{\bf S}_{(L)}
=
\left(
\baa{cccc}
e^{- 2 \tilde\gamma_0 \tau }& 0 & 0 & 0\\
 0 & f_{1}  (\tau) e^{- 3 \tilde\gamma_0 \tau }    &
\frac{\sin{(\kappa \tau)} }{\kappa} e^{- 3 \tilde\gamma_0 \tau } & 
\frac{4 \tilde\gamma_0}{8 \tilde\gamma_0^2 +1} \left(f_{3}  (\tau) e^{- 3 \tilde\gamma_0 \tau }   -1\right) \\
 0 & -\frac{\sin{(\kappa \tau)} }{\kappa} e^{- 3 \tilde\gamma_0 \tau } & f_{-1}  (\tau) e^{- 3 \tilde\gamma_0 \tau }  & 
\frac{8 \tilde\gamma_0^2}{8 \tilde\gamma_0^2 +1} \left(f_{- \nu^2}  (\tau) e^{- 3 \tilde\gamma_0 \tau }   -1\right)\\
 0 & 0 & 0 & 1 
\eaa
\right)
,\nn
\ea
\ew
where
$f_{k} (\tau) \equiv \cos{(\kappa \tau)} + k \frac{\tilde\gamma_0 }{\kappa}\sin{(\kappa \tau)} $ and
$\nu^2 = 1 + 1/(2 \tilde\gamma_0^2)$.
From the last row of this matrix it follows that 
$\langle \sigma_4 (\tau) \rangle_{(I)} = {\rm tr} (\hat\rho (\tau)) = 1$.
Therefore, the physical (normalized) averages
coincide with the auxiliary ones: $\langle \sigma_i' (t) \rangle = \langle \sigma_i (t)\rangle$, $i=1..3$.
It is easy to check that this solution coincides with a textbook example - see, for instance, section
3.4.5.1 of \cite{bpbook}.
The profiles of most important observables are given in the Figs. \ref{f-bp-pe} 
and \ref{f-bp-sigy}.

The large-times asymptotic (steady-state) values of the spin averages can be found by 
taking an appropriate limit in equation (\ref{eq:gensolL}).
One obtains
\be
\langle \vec\sigma' (+ \infty) \rangle_{(I)}
=
\langle \vec\sigma (+ \infty) \rangle_{(I)}
=
-
\frac{4 \tilde\gamma_0}{8 \tilde\gamma_0^2 +1}
\left(
\begin{array}{c} 
0 \\
1 \\
2 \tilde\gamma_0
\end{array}
\right)
.
\ee
These values are indeed stationary points of the system.
Needless to say, they coincide with the textbook values, cf. section
3.4.5.1 of \cite{bpbook}.

\subsection{Anti-Hermitian-driven dissipation}\lb{s-lc-ah}

When $\tilde\gamma_0 =0$ then the matrix $\widetilde{\bf M}$
has the following four eigenvalues:
\bw
\begin{eqnarray}
&&
(M_1^{(A)},M_2^{(A)},M_3^{(A)},M_4^{(A)})
=
(\lambda_1, \,\lambda_2, \,\lambda_3, \,\lambda_4) =
\nonumber\\
&& \qquad =
\left(
- \sqrt{\frac{R_+ - R_1}{2}}, \,
\sqrt{\frac{R_+ - R_1}{2}}, \,
- \sqrt{\frac{R_+ + R_1}{2}}, \,
\sqrt{\frac{R_+ + R_1}{2}},
\right),
\lb{eq:evaAH}
\end{eqnarray} 
where we have denoted
$
R_1 = 
\sqrt{
R_+^2 + 4 \tilde\alpha^2
}
=
\sqrt{
R_-^2 + 4 \tilde\alpha^2 \tilde\Gamma^2
}
$ and $R_\pm =
\tilde\alpha^2 \pm (\tilde\Gamma^2 -1)
.
$
One can see that $R_1 \geqslant R_\pm$ so that the first two eigenvalues (\ref{eq:evaAH})
are always imaginary-valued.
Correspondingly, the solution from Sec. \ref{s-gsol}
can be written as
\ba
&&
\langle \sigma_\mu (\tau) \rangle_{(I)}
=
%|| \text{diag} (\exp{M_{(1)} \tau}) ||
{\bf S}_{(A)}
\langle \sigma_\mu (0) \rangle_{(I)}
,
\lb{eq:gensolAH}\\&&
{\bf S}_{(A)}
=
\frac{1}{R_1}
\left(
\baa{cccc}
\frac{1}{2} C_{K_-, K_+} (\tau)& 
\tilde\alpha \tilde\Gamma  S_{-\lambda_2^{-1}, \lambda_4^{-1}} (\tau) & 
\tilde\alpha \tilde\Gamma  C_{1, -1} (\tau) & 
\tilde\alpha  S_{\bar\lambda_2, -\bar\lambda_4} (\tau)\\
\frac{\lambda_2 \lambda_4 \tilde\Gamma }{\tilde\alpha } S_{-\lambda_4, \lambda_2} (\tau)& 
C_{\lambda_4 \bar\lambda_4, -\lambda_2 \bar\lambda_2} (\tau) & 
- \lambda_2 \lambda_4  S_{\bar\lambda_4 , -\bar\lambda_2} (\tau) & 
- \tilde\Gamma  C_{1, -1} (\tau)\\
\tilde\alpha \tilde\Gamma  C_{1, -1} (\tau) & 
\lambda_2 \lambda_4  S_{\bar\lambda_4 , -\bar\lambda_2} (\tau) & 
\frac{1}{2} C_{K_+, K_-} (\tau)& 
\tilde\Gamma  S_{-\lambda_2, \lambda_4} (\tau)\\
-\frac{1}{2 \tilde\alpha} S_{\lambda_2 K_-, \lambda_4 K_+} (\tau)& 
\tilde\Gamma  C_{1, -1} (\tau) & 
\tilde\Gamma  S_{-\lambda_2, \lambda_4} (\tau)&
C_{-\lambda_2 \bar\lambda_2, \lambda_4 \bar\lambda_4} (\tau)
\eaa
\right)
,
\nn
\ea
where we have denoted
\ba
&&
C_{k_1,k_2}(\tau) \equiv k_1 \cosh{(\lambda_2 \tau)} +  k_2 \cosh{(\lambda_4 \tau)} = 
k_1 \cos{(|\lambda_2| \tau)} +  k_2 \cosh{(|\lambda_4 | \tau)}
,\nn\\&&
S_{k_1,k_2}(\tau) \equiv k_1 \sinh{(\lambda_2 \tau)} +  k_2 \sinh{(\lambda_4 \tau)} = 
i k_1 \sin{(|\lambda_2| \tau)} +  k_2 \sinh{(|\lambda_4 |\tau)}
,\nn
\ea
and
$K_\pm = R_1 \pm R_- = \sqrt{
R_-^2 + 4 \tilde\alpha^2 \tilde\Gamma^2
} \pm R_-$
and
$\bar\lambda_k = \lambda_k + 1/\lambda_k$.
As before, the physical values are the normalized ones:
\ba
&&
\langle \vec\sigma' (\tau) \rangle_{(I)}
=
%|| \text{diag} (\exp{M_{(1)} \tau}) ||
{\bf S}_{(A)}'
\langle \sigma_\mu (0) \rangle_{(I)}
,
\lb{eq:gensolAHn}\\&&
{\bf S}_{(A)}'
=
\frac{1}{T_{(A)}}
\left(
\baa{cccc}
\frac{1}{2} C_{K_-, K_+} (\tau)& 
\tilde\alpha \tilde\Gamma  S_{-\lambda_2^{-1}, \lambda_4^{-1}} (\tau) & 
\tilde\alpha \tilde\Gamma  C_{1, -1} (\tau) & 
\tilde\alpha  S_{\bar\lambda_2, -\bar\lambda_4} (\tau)\\
\frac{\lambda_2 \lambda_4 \tilde\Gamma }{\tilde\alpha } S_{-\lambda_4, \lambda_2} (\tau)& 
C_{\lambda_4 \bar\lambda_4, -\lambda_2 \bar\lambda_2} (\tau) & 
- \lambda_2 \lambda_4  S_{\bar\lambda_4 , -\bar\lambda_2} (\tau) & 
- \tilde\Gamma  C_{1, -1} (\tau)\\
\tilde\alpha \tilde\Gamma  C_{1, -1} (\tau) & 
\lambda_2 \lambda_4  S_{\bar\lambda_4 , -\bar\lambda_2} (\tau) & 
\frac{1}{2} C_{K_+, K_-} (\tau)& 
\tilde\Gamma  S_{-\lambda_2, \lambda_4} (\tau)
%\\ -\frac{1}{2 \tilde\alpha} S_{\lambda_2 K_-, \lambda_4 K_+} (\tau)& \tilde\Gamma  C_{1, -1} (\tau) & 
%\tilde\Gamma  S_{-\lambda_2, \lambda_4} (\tau)& C_{-\lambda_2 \bar\lambda_2, \lambda_4 \bar\lambda_4} (\tau)
\eaa
\right)
,
\nn
\ea
where
$T_{(A)}$ is an internal product of the fourth row of ${\bf S}_{(A)}$ 
(omitting the overall factor)
and the four-vector of 
initial values
(\ref{eq:ini4values}):
\[
T_{(A)} =
-\frac{1}{2 \tilde\alpha} S_{\lambda_2 K_-, \lambda_4 K_+} (\tau)
\langle \sigma_1 (0) \rangle
+
\tilde\Gamma  C_{1, -1} (\tau) 
\langle \sigma_2 (0) \rangle
+
\tilde\Gamma  S_{-\lambda_2, \lambda_4} (\tau)
\langle \sigma_3 (0) \rangle
+
C_{-\lambda_2 \bar\lambda_2, \lambda_4 \bar\lambda_4} (\tau)
.
\]
\ew
Depending on whether the eigenvalue $\lambda_4$ vanishes or not, one can consider
the following two cases.

\begin{figure}[htbt]
\begin{center}\epsfig{figure=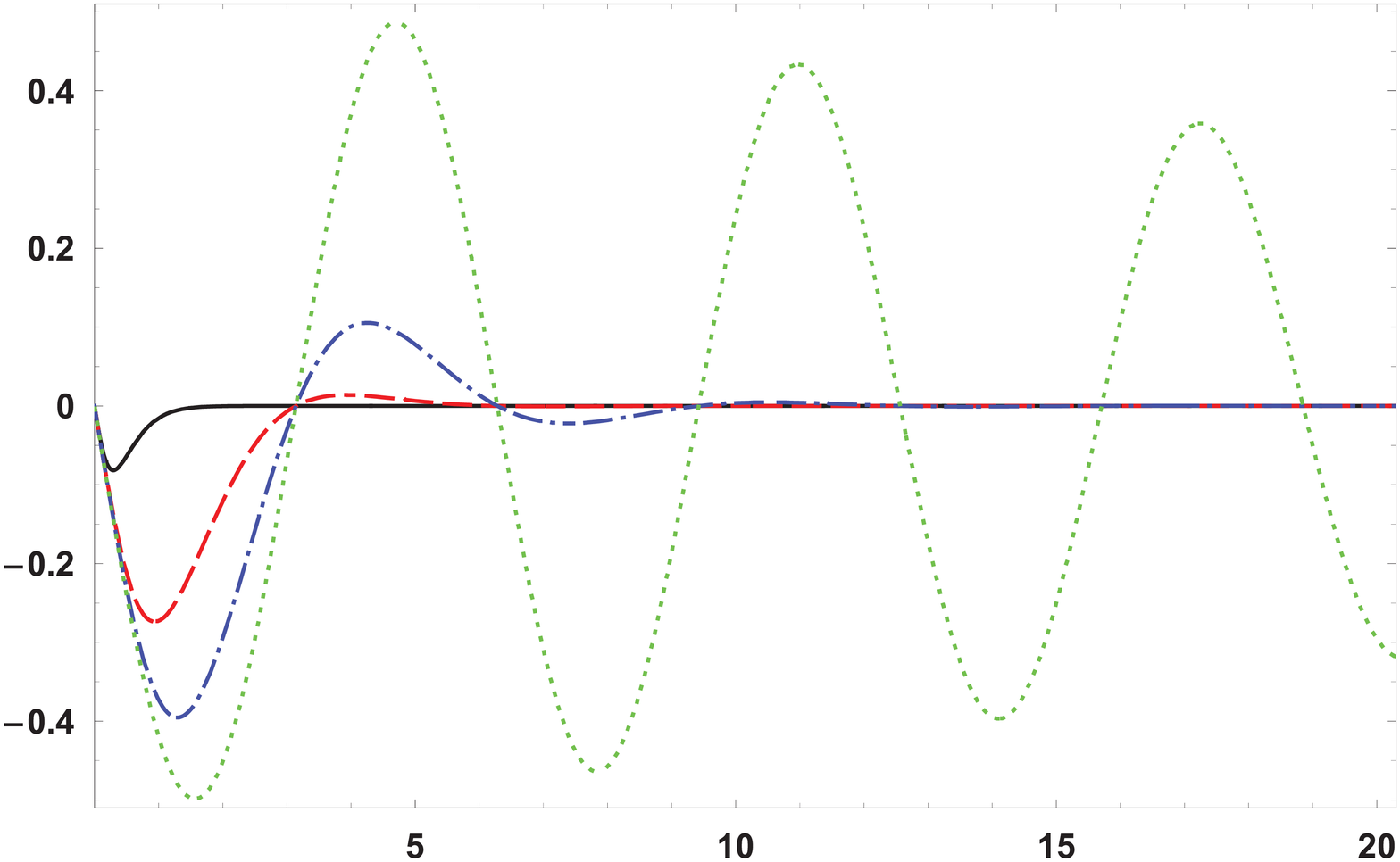,width=1.01\columnwidth}\end{center}
\caption{
The imaginary part of the coherence, $\text{Im} \left(\left\langle {\sigma}_+ \right\rangle_\text{obs} \right)$, 
as a function of time 
$\tau = \Omega t$ for the parameters $\tilde\gamma_0 =\tilde\Gamma = 0$ and:
$\tilde\alpha = 4$ (solid curve),
$\tilde\alpha = 1$ (dashed curve),
$\tilde\alpha = 1/2$ (dash-dotted curve),
and $\tilde\alpha = 1/20$ (dotted curve).}
\label{f-g0-sigy}
\end{figure}

\begin{figure}[htbt]
\begin{center}\epsfig{figure=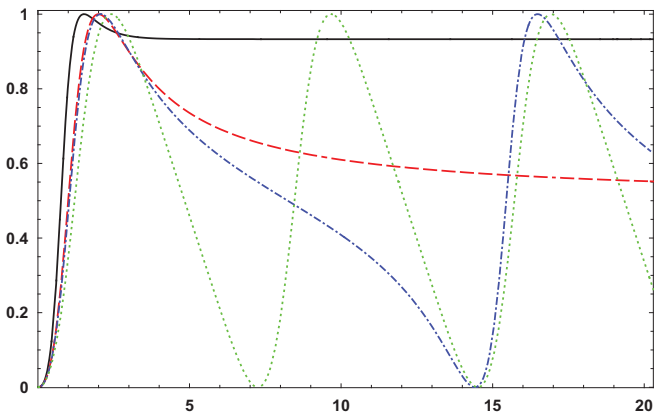,width=1.01\columnwidth}\end{center}
\caption{
The population of the upper level $p_e$ as a function of time 
$\tau = \Omega t$ for the parameters $\tilde\alpha = \tilde\gamma_0 =0$ and:
$\tilde\Gamma = 2$ (solid curve),
$\tilde\Gamma = 1$ (dashed curve),
$\tilde\Gamma = 0.9$ (dash-dotted curve),
and $\tilde\Gamma = 1/2$ (dotted curve).}
\label{f-a0-pe}
\end{figure}

\subsubsection{Exponential damping}

This case takes place when the parameters of the model are such that
$
\lambda_4 \not= 0
,
$
or, alternatively,
$ 
%|R_+| \left(
\Sign{R_+} + \sqrt{
1 + (2 \tilde\alpha/ R_+)^2 } 
%\right)
\not = 0$. Then the system exhibits the exponential damping which is qualitatively, but not necessarily quantitatively, similar
to the Lindblad-driven dynamics.

Some generic profiles of most important observables are given 
in Figs. \ref{f-g0-pe} and \ref{f-g0-sigy} (all curves), 
as well as in Figs.  \ref{f-a0-pe} and \ref{f-a0-sigy}
(solid and dashed curves only).
One can notice that Figs.~\ref{f-g0-pe} and \ref{f-g0-sigy} qualitatively resemble
the Lindblad ones.

The large-times asymptotic values of the spin averages can be found by taking an appropriate limit in (\ref{eq:gensolAHn}).
At first one obtains
\bw
\[
\langle \vec\sigma' (+ \infty) \rangle_{(I)}
=
\frac{1}{\tilde T_{(A)}}
\left(
\baa{cccc}
\frac{1}{2} \tilde\alpha  K_+& 
\tilde\alpha^2   \lambda_4^{-1} \tilde\Gamma & 
-\tilde\alpha^2 \tilde\Gamma  & 
\tilde\alpha^2  \bar\lambda_4\\
\lambda_2^2 \lambda_4 \tilde\Gamma  & 
-\tilde\alpha  \lambda_2 \bar\lambda_2 & 
\tilde\alpha  \lambda_2 \bar\lambda_2 \lambda_4   & 
\tilde\alpha  \tilde\Gamma  \\
-\tilde\alpha^2 \tilde\Gamma  & 
- \tilde\alpha  \lambda_2 \bar\lambda_2 \lambda_4 & 
\frac{1}{2} \tilde\alpha  K_- & 
\tilde\alpha \lambda_4 \tilde\Gamma  
\eaa
\right)
\langle \vec\sigma  (0) \rangle
,
\]
\ew
where we have denoted
$
\tilde T_{(A)} =
-\frac{1}{2 } \lambda_4 K_+
\langle \sigma_1 (0) \rangle
-
\tilde\alpha 
\tilde\Gamma   
\langle \sigma_2 (0) \rangle
+
\tilde\alpha \lambda_4
\tilde\Gamma  
\langle \sigma_3 (0) \rangle
+
\tilde\alpha \lambda_4 \bar\lambda_4
.
$
Further simplifying this expression,
we eventually obtain
\be
\langle \vec\sigma' (+ \infty) \rangle_{(I)}
=
%\langle \vec\sigma (+ \infty) \rangle_{(I)} =
%\frac{4 \tilde\gamma_0}{8 \tilde\gamma_0^2 +1}
\frac{1}{\lambda_4 \bar\lambda_4}
\left(
\begin{array}{c} 
- \tilde\alpha \bar\lambda_4  \\
\tilde\Gamma  \\
\lambda_4  \tilde\Gamma 
\end{array}
\right)
,
\ee
which means that the asymptotic (steady-state) averages do not depend
on the initial values, as in the Lindblad case.
One can see that the important difference from the Lindblad case is that
$\langle \sigma'_x \rangle_{(I)}$ does not vanish at large times.

Another distinctive features of the anti-Hermitian-driven dynamics can be
found if one computes the Fourier transform of the observables, such as the
population of an upper level $p_e$.
The informative part of the Fourier transform is,
according to (\ref{e:ft-mon}),
\be\lb{e:pe-ft-ah-dec}
[p_e (\tilde\omega)]_\text{reg} 
\propto
\frac{1}{\sqrt{2 \pi}}
\int\limits_{0}^\infty
\left( p_e (\tau) - p_e(\infty) \right) \text{e}^{i \tilde\omega \tau} d \tau
,
\ee
where $\tilde\omega = \omega / \Omega$, and its functional dependence can be derived with the use of the equations (\ref{eq:gensolAHn}),
(\ref{e:pe-sig3}) and (\ref{e:intpic-vecsig}).
Since the Fourier transforms are essentially complex-valued,
one should consider separately moduli and phases.

\begin{figure}[htbt]
\begin{center}\epsfig{figure=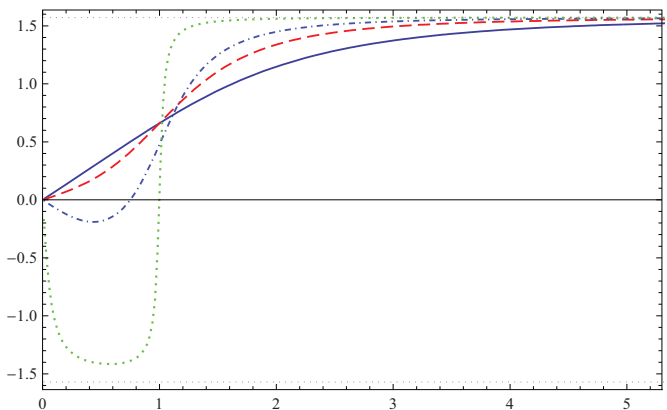,width=1.01\columnwidth}\end{center}
\caption{
The phase of the Fourier transform $p_e (\omega)$ 
%(\ref{e:pe-ft-ah-dec})
versus the frequency $\tilde\omega = \omega / \Omega$,
for the parameters $\tilde\alpha =\tilde\Gamma = 0$ and:
$\tilde\gamma_0 = 1/2$ (solid curve),
$\tilde\gamma_0= 1/4$ (dashed curve),
$\tilde\gamma_0 = 1/8$ (dash-dotted curve),
and $\tilde\gamma_0 = 1/100$ (dotted curve).
Two horizontal thin dotted lines mark the values $\pm \pi/2$. }
\label{f-bp-pe-ftarg}
\end{figure}

\begin{figure}[htbt]
\begin{center}\epsfig{figure=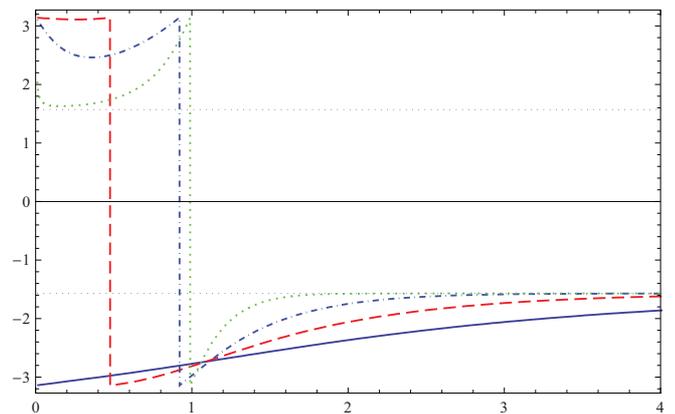,width=1.01\columnwidth}\end{center}
\caption{
The phase of the Fourier transform $p_e (\omega)$ (\ref{e:pe-ft-ah-dec})
versus the frequency $\tilde\omega = \omega / \Omega$,
for the parameters $\tilde\gamma_0 =\tilde\Gamma = 0$ and:
$\tilde\alpha = 2$ (solid curve),
$\tilde\alpha = 1$ (dashed curve),
$\tilde\alpha = 1/2$ (dash-dotted curve),
and $\tilde\alpha = 1/5$ (dotted curve).
Two horizontal thin dotted lines mark the values $\pm \pi/2$. }
\label{f-g0-pe-ftarg}
\end{figure}

It turns out that the absolute value of $[p_e (\tilde\omega)]_\text{reg} $ behaves 
in a qualitative similar way to its Lindblad-driven counterpart, 
under the same initial conditions.
The differences arise when one consider the phase of $[p_e (\tilde\omega)]_\text{reg} $.
While in the Lindblad-driven case it is a smooth function for positive $\tilde\omega$
which is bound between $-\pi/2$ and $\pi/2$ 
with the asymptotic value $\pi /2$ (see figure \ref{f-bp-pe-ftarg}),
in the anti-Hermitian-driven case it is exactly opposite.
From the figure \ref{f-g0-pe-ftarg} 
one can see that
the phase of $[p_e (\tilde\omega)]_\text{reg}$ varies between $\pi/2$ and $-\pi/2$ 
with the asymptotic value $-\pi /2$.

\bw
\subsubsection{Anharmonic oscillations}

In this case, the model parameters are such that
$
\lambda_4 = 0
,
$
which is equivalent to the following two conditions:
$
\tilde\alpha = 0
$ and
$
\tilde\Gamma^2  < 1
.
$
The solution exhibits purely oscillatory behavior:
\be
\langle \vec\sigma' (\tau) \rangle_{(I)}
=
\frac{1}{T_{(A)}^{\Gamma}}
\left(
\begin{array}{c} 
\omega_\Gamma^2 \langle \sigma_1 (0) \rangle \\
\tilde\Gamma - \tilde\Gamma^2 \langle \sigma_2 (0) \rangle
- 
%D\!\left(, \ ; \tau \right)
(\tilde\Gamma - \langle \sigma_2 (0) \rangle) \cos{(\omega_\Gamma \tau)} +  \omega_\Gamma \langle \sigma_3 (0) \rangle \sin{(\omega_\Gamma \tau)}
\\
%\omega_\Gamma 
%D\!\left(\omega_\Gamma \langle \sigma_3 (0) \rangle , \ \tilde\Gamma  - \langle \sigma_2 (0) \rangle; \tau \right) 
\omega_\Gamma^2 \langle \sigma_3 (0) \rangle \cos{(\omega_\Gamma \tau)} + (\tilde\Gamma  - \langle \sigma_2 (0) \rangle) \sin{(\omega_\Gamma \tau)}
\end{array}
\right)
,
\lb{e:gensolAHosc}
\ee
\ew
where the oscillation frequency $\omega_\Gamma = \sqrt{1 - \tilde\Gamma^2}$, and
$
T_{(A)}^{\Gamma} =
1- \tilde\Gamma \langle \sigma_2 (0) \rangle
- 
\tilde\Gamma 
%D\!\left( \langle \sigma_2 (0) \rangle - \tilde\Gamma, \ \omega_\Gamma \langle \sigma_3 (0) \rangle; \tau \right)
(\tilde\Gamma - \langle \sigma_2 (0) \rangle) \cos{(\omega_\Gamma \tau)} +  \omega_\Gamma \langle \sigma_3 (0) \rangle \sin{(\omega_\Gamma \tau)}
.
$
%$ D (k_1,  k_2; \tau) \equiv k_1 \cos{(\omega_\Gamma \tau)} +  k_2 \sin{(\omega_\Gamma \tau)},$ .

\begin{figure}[htbt]
\begin{center}\epsfig{figure=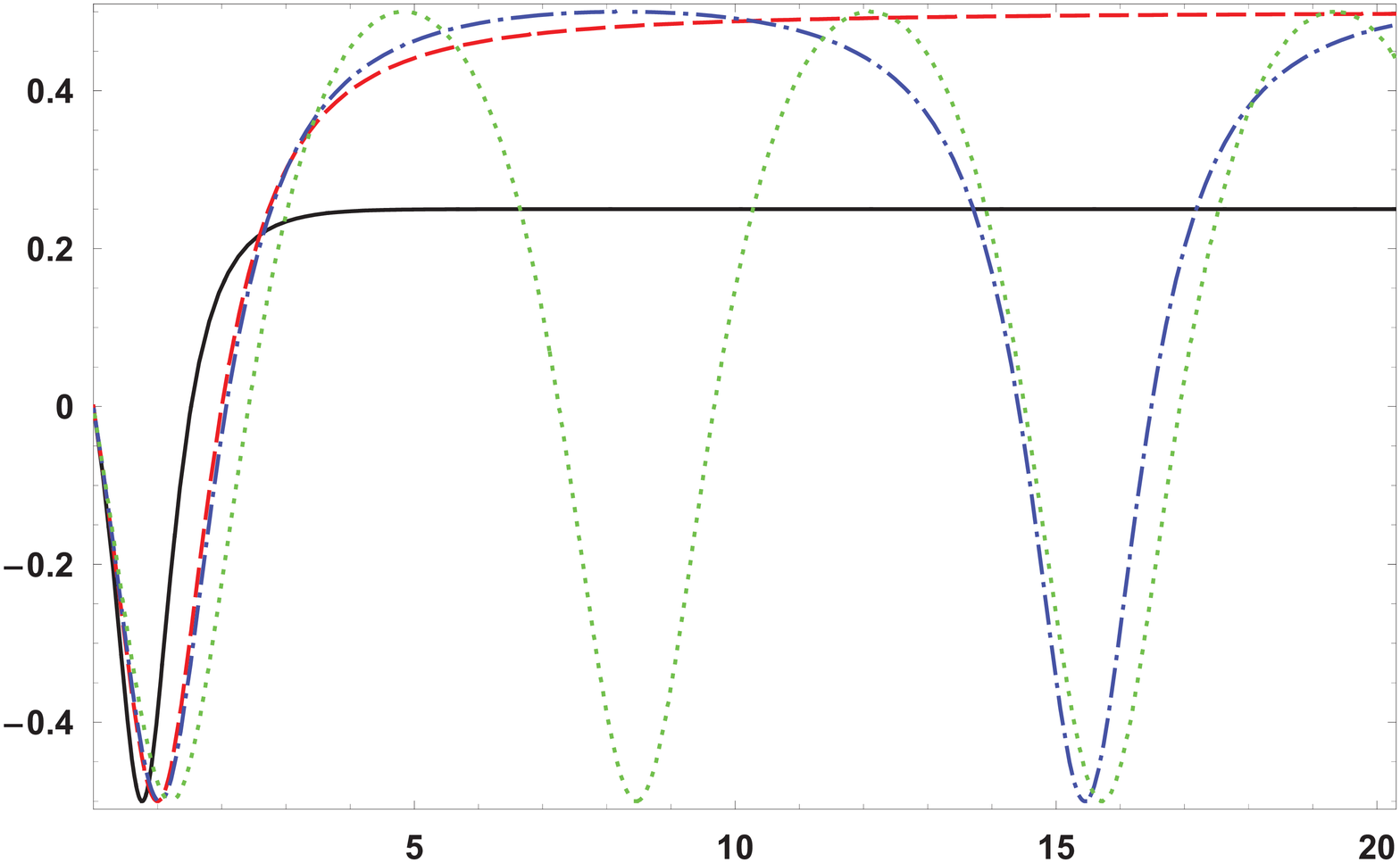,width=1.01\columnwidth}\end{center}
\caption{
The imaginary part of the coherence, $\text{Im} \left(\left\langle {\sigma}_+ \right\rangle_\text{obs} \right)$, 
as a function of time 
$\tau = \Omega t$ for the parameters $\tilde\alpha = \tilde\gamma_0 =0$ and:
$\tilde\Gamma = 2$ (solid curve),
$\tilde\Gamma = 1$ (dashed curve),
$\tilde\Gamma = 0.9$ (dash-dotted curve),
and $\tilde\Gamma = 1/2$ (dotted curve).}
\label{f-a0-sigy}
\end{figure}

\begin{figure}[htbt]
\begin{center}\epsfig{figure=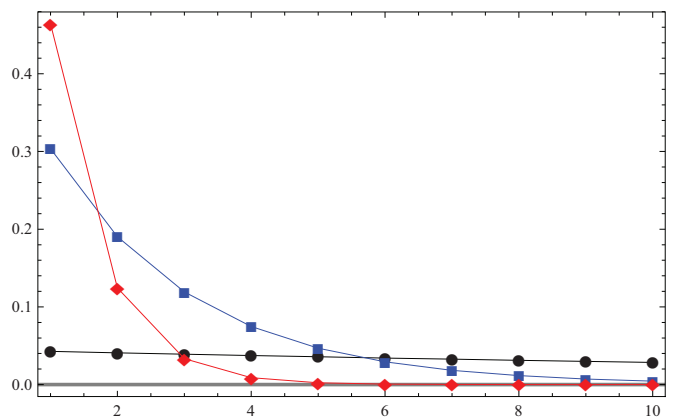,width=1.01\columnwidth}\end{center}
\caption{
The modulus of the Fourier transform $p_e (n)$ (\ref{e:pe-ft-ah-osc})
versus the wavenumber $n$,
for the parameters $\tilde\alpha = \tilde\gamma_0 =0$ and:
$\tilde\Gamma = 0.999$ (dots),
$\tilde\Gamma = 0.9$ (squares),
and $\tilde\Gamma = 1/2$ (diamonds).}
\label{f-a0-pe-ftabs}
\end{figure}

Some profiles of most important observables are given in Figs. \ref{f-a0-pe} 
and \ref{f-a0-sigy} (dotted and dash-dotted curves only).
They exhibit interesting
asymmetric oscillatory patterns which do not appear in the Lindblad case.
Such patterns indicate a few important things.
For instance, they show that the anti-Hermitian terms in the Hamiltonian can induce not only 
the standard decay effects (such as the asymptotic damping at large times) 
but also more sophisticated effects.
Indeed, in this case the oscillations are not damped, the role of the anti-Hermitian parameter
$\tilde\Gamma$ is that it introduces the asymmetry between the pumping and discharging of the two-level system.
In terms of frequency it means that the pumping frequency is larger than the discharge one.
It is similar to what happens in higher-than-two-level systems:
first a system is pumped
into the highest excited state, then it spontaneously cascades down to its ground state, 
passing the intermediate levels on its way.
This is particularly clear to see when one takes a look at the Fourier transform of the
population of the upper level $p_e$.
The informative part of the Fourier transform is,
according to (\ref{e:ft-osc}),
\be\lb{e:pe-ft-ah-osc}
[p_e (n)]_\text{reg} 
\propto
\frac{1}{T_o}
\int\limits_{0}^{T_o}
p_e (\tau)  \text{e}^{2 \pi i (n / T_o) \tau} d \tau
,
\ee
where $T_o = 2 \pi/ \sqrt{1 - \tilde\Gamma^2}$, 
and its specific functional form can be derived with the use of the equations (\ref{e:gensolAHosc}),
(\ref{e:pe-sig3}) and (\ref{e:intpic-vecsig}). 
The profile of the computed modulus of $[p_e (n)]_\text{reg}$ is shown in the 
figure \ref{f-a0-pe-ftabs}.

To summarize, we have shown that the anti-Hermitian two-level models of this type can actually
mimic the properties of quantum systems with more than two levels.
In this picture the parameter $\tilde\Gamma$ turns out to be a qualitative measure of the number of the additional (effective) levels. 
For example, the figure \ref{f-a0-pe-ftabs} shows that
the TLS with $\tilde\Gamma = 1/2$ can be used to mimic the 4-level or 5-level system 
(if one neglects the ``transitions'' with the wavenumbers larger than four).
The further decreasing of $\tilde\Gamma $ reduces the number of additional wave frequencies.

\section{Approximate solution}\lb{s-apsol}

The general solution derived in Sec. \ref{s-gsol} becomes extremely bulky
when expressed in terms of radicals.
Luckily, in some physical cases one could use certain approximations which drastically
simplify final formulae.
Indeed, in quantum-optical two-level systems the Rabi frequency usually takes large
values, up to the megahertz scale, whereas the dissipative effects are small.
It is thus natural to make the assumption
\be\lb{e:smalltilde}
\tilde\gamma_0 \ll 1,\
\tilde\alpha \ll 1,\
\tilde\Gamma \ll 1,
\ee
which corresponds to the strong driving limit, using the 
textbook terminology \cite{bpbook}.
However, this condition is not enough since the perturbation theory has
two different sectors - Lindblad-dominated (when the approximate series solution must converge to the solution
from Sec. \ref{s-lc-li} when taking the limit $\alpha = \Gamma =0$) and anti-Hermitian-dominated
(when the series solution must converge to the solution
from Sec. \ref{s-lc-ah} in the limit $\gamma =0$).
In the former case, which is of main interest here, one should supplement (\ref{e:smalltilde})
with the assumption
\be
\bar\alpha \equiv \alpha/\gamma_0 \ll 1,\
\bar\Gamma \equiv \Gamma /\gamma_0  \ll 1,
\ee
then
expand the exact solution (\ref{eq:gensolB}) and the related observables in series with respect to 
these five small parameters, and keep the leading-order terms.
By doing that we obtain that the matrix $\widetilde{\bf M}$
has the following four eigenvalues, in the leading-order approximation:
\bw
\be
(M_1^{(LD)},M_2^{(LD)},M_3^{(LD)},M_4^{(LD)})
\approx
(- 2 \tilde\gamma_0 - 2 \tilde\gamma_0 \chi_2  , \,
- 3 \tilde\gamma_0 - i \chi_1, \,
- 3 \tilde\gamma_0 + i \chi_1 , \,
2 \tilde\gamma_0 \chi_2 ),
\ee 
where we denoted
$\chi_1 = 1 + (2 \bar\Gamma - 1/2) \tilde\gamma_0^2$
and
$\chi_2 = \bar\alpha^2/4$, both being positive-definite values.
Correspondingly,
the solution is given by
\ba
&&
\langle \sigma_\mu (\tau) \rangle_{(I)}
=
%|| \text{diag} (\exp{M_{(1)} \tau}) ||
{\bf S}_{(LD)}
\langle \sigma_\mu (0) \rangle_{(I)}
,
\lb{eq:gensolAL}
\\&&
{\bf S}_{(LD)}
\approx
{\bf S}_{(LD)}^{(0)}
+
{\bf S}_{(LD)}^{(\alpha)}
+
{\bf S}_{(LD)}^{(\Gamma)}
,\nn
\ea
\ba
&&
{\bf S}_{(LD)}^{(0)}
=
\left(
\baa{cccc}
e^{- 2 \tilde\gamma_0 (1+ \chi_2)\tau }
& 
0
& 0 & 0
\\
0 
& 
h_{+}  (\tau) e^{- 3 \tilde\gamma_0 \tau }    
&
\sin{(\chi_1 \tau)} e^{- 3 \tilde\gamma_0 \tau } 
& 
4 \tilde\gamma_0 h_2 (\tau) 
\\
0 
& 
- \sin{(\chi_1 \tau)} e^{- 3 \tilde\gamma_0 \tau }
& 
h_{-}  (\tau) e^{- 3 \tilde\gamma_0 \tau }  
& 
- 4 \tilde\gamma_0 \sin{(\chi_1 \tau)} e^{- 3 \tilde\gamma_0 \tau }
\\
 0 & 0 & 0 
& 
e^{2 \tilde\gamma_0  \chi_2 \tau }
\eaa
\right)
,
\nn\\&&
{\bf S}_{(LD)}^{(\alpha)}
=
\bar\alpha 
%h_1 (\tau)
\left(
\baa{cccc}
0
& 
0
& 0 & 
h_1 (\tau)
\\
- 4 \tilde\gamma_0 h_1 (\tau)
& 
0
&
0 
& 
0 
\\
- 4 \tilde\gamma_0^2 h_2 (\tau) 
& 
0
& 
0 
& 
0
\\
h_1 (\tau)
& 0 & 0 
& 
0
\eaa
\right)
, 
\nn\\&&
{\bf S}_{(LD)}^{(\Gamma)}
=
%\tilde\gamma_0 \bar\Gamma
\tilde\Gamma
\left(
\baa{cccc}
0
& 
- \bar\alpha h_1 (\tau)
& 
\tilde\alpha h_2 (\tau) 
& 
0
\\
\bar\alpha h_1 (\tau)
& 
0
&
0 
& 
- h_2 (\tau) 
\\
\tilde\alpha h_2 (\tau)
& 
0
& 
0 
& 
\sin{(\chi_1 \tau)} e^{- 3 \tilde\gamma_0 \tau }
\\
0
& 
h_2 (\tau) 
& 
\sin{(\chi_1 \tau)} e^{- 3 \tilde\gamma_0 \tau }
& 
0
\eaa
\right)
,
\ea
where
$
h_1 (\tau) = 
\tfrac{1}{2}
\left(
e^{- 2 \tilde\gamma_0 (1+ \chi_2)\tau } 
-
e^{2 \tilde\gamma_0  \chi_2 \tau }
\right)
$,
$
h_2 (\tau) = 
\cos{(\chi_1 \tau)}
e^{- 3 \tilde\gamma_0 \tau }  
-
e^{2 \tilde\gamma_0  \chi_2 \tau }
$,
and
$
h_\pm (\tau) = \cos{(\chi_1 \tau)} \pm \tilde\gamma_0 \sin{(\chi_1 \tau)} 
$.
As in Sec. \ref{s-lc-ah}, the physical values are the normalized ones:
\be\lb{eq:gensolALn}
\langle \vec\sigma' (\tau) \rangle_{(I)}
=
%|| \text{diag} (\exp{M_{(1)} \tau}) ||
\frac{1}{T_{(LD)}}
{\bf S}_{(LD)}'
\langle \sigma_\mu (0) \rangle_{(I)}
,
\ee
where ${\bf S}_{(LD)}'$ is the matrix ${\bf S}_{(LD)}$ without the bottom row,
$T_{(LD)}$ is a product of the bottom row of ${\bf S}_{(LD)}$ 
%(omitting the overall factor)
and the four-vector of 
initial values
(\ref{eq:ini4values}):
\[
T_{(LD)} =
\bar\alpha h_1 (\tau)
\langle \sigma_1 (0) \rangle
+
\tilde\Gamma h_2 (\tau)
\langle \sigma_2 (0) \rangle
+
\tilde\Gamma
\sin{(\chi_1 \tau)} e^{- 3 \tilde\gamma_0 \tau }
\langle \sigma_3 (0) \rangle
+
e^{2 \tilde\gamma_0  \chi_2 \tau }
.
\] 
\ew
The large-times asymptotic values of the spin averages
can be found by taking an appropriate limit in (\ref{eq:gensolALn}).
Hence, we obtain in the leading-order approximation
\be
\langle \vec\sigma' (+ \infty) \rangle_{(I)}
=
\left(
\begin{array}{c} 
- \frac{1}{2} \bar\alpha  - 2 \tilde\alpha \tilde\Gamma\\
\tilde\Gamma - 4 \tilde\gamma_0\\
2 \tilde\gamma_0 (\tilde\Gamma - 4 \tilde\gamma_0)
\end{array}
\right)
%+ {\cal O} (\tilde\gamma_0^k \tilde\alpha^m \tilde\Gamma^n; k+m+n > 1)
.
\ee

\section{Conclusion}\lb{s-con}

In this paper 
we have compared two approaches to describing the effects of a general dissipative environment.
Namely, we considered both the approach based on the Lindblad master equation
and the formalism based on introducing anti-Hermitian terms into the Hamiltonian.
In section \ref{s-uni} we have proposed a ``hybrid'' formalism
that unifies the Lindblad and non-Hermitian approaches.
This allowed us not only to reveal the distinctive features
of the approaches but also to expand the range of dissipative phenomena that can
be accounted for.

Using a two-level single-atom model as a practical application,
we have obtained solutions of the hybrid equation
for the normalized density matrix operator.
In sections \ref{s-gsol}, \ref{s-lc} and \ref{s-apsol},
we have also considered special (limit) cases and physically admissible approximations.
Using the analytical solutions of all these cases, 
we have calculated those properties of the model
that can be compared with experiments
in order to assess whether a specific feature is either non-Hermitian driven
or Lindblad driven.

Remarkably,
we have found that the anti-Hermitian terms in the Hamiltonian can describe not only the
mere dissipative damping but also undamped anharmonic oscillatory phenomena.
Such results are reported in detail in section \ref{s-lc-ah}
where we also showed that this kind of anharmonicity can be used to mimic 
the cascaded quantum systems with more than two levels.
In future it would be interesting to apply the hybrid formalism
to those multi-level lasers or spin systems that can be modeled  in the leading order of approximation
by means of only two states.
In particular, we have in mind those systems 
where one of the two energy levels
is actually a band or a bundle of a few closely situated levels.

\section*{Acknowledgments}

This article is based on the talks given at the conferences
``12th International Workshop on Pseudo-Hermitian Hamiltonians in Quantum Physics'' 
(02-06 July, 2013, Ko\c{c} University, Istanbul, Turkey)
and
``Quantum Information Processing, Communication and Control 2''
(25-29 November, 2013, KwaZulu-Natal, South Africa).
This work was supported by
the National Research Foundation of South Africa.

%\newpage
\appendix

\section{Two-level systems in quantum optics}\lb{s-a-tls}

The two-mode open quantum system is a basic yet very instructive example of an open quantum (sub)system.
In quantum optics its most obvious manifestation is the two-level atom interacting
with the external electromagnetic field (such as the laser field)
and dissipative environment (heat bath, noise, etc).
However, two-level models can also serve as a decent first-order approximation for those physical phenomena
whose dynamics is effectively confined to a two-dimensional subspace,
one example to be the systems for which one can neglect the influence of excited levels above the first excited one.
%,or where the excited states are densely distributed compared to the ground one.
Here we outline the basic notions used in a theory of two-level quantum optical systems.

For a general two-level quantum system the Hilbert space has the dimensionality two,
and 
it is spanned by just two states, a ground state $|g\rangle$ and an excited state $|e\rangle$.
An arbitrary quantum state of such system can be written in the basis of the Pauli and unit matrices
which form a complete set.
In quantum optics one is often interested in such averaged values
as 
the \textit{population difference}
\be
\left\langle {\sigma}_3 \right\rangle_\text{obs}
\equiv
{\rm tr}\left(\hat{\rho}\hat{\sigma}_3 \right)/ {\rm tr}\left(\hat{\rho}\right) =
\frac{
\rho_{11}  - \rho_{22} 
}{
\rho_{11} + \rho_{22}
}
,
\ee
the \textit{population of the excited-state (upper) level}
\be\lb{e:pe-sig3}
p_e
%\equiv {\rm tr}\left(\hat{\rho}\hat{\sigma}_3 \right)/ {\rm tr}\left(\hat{\rho}\right) 
=
\frac{
\rho_{11} 
}{
\rho_{11} + \rho_{22}
}
=
\frac{1}{2}
\left(
1 +
\left\langle {\sigma}_3 \right\rangle_\text{obs}
\right)
,
\ee
the \textit{population of the ground state level}
\be
p_g
%\equiv {\rm tr}\left(\hat{\rho}\hat{\sigma}_3 \right)/ {\rm tr}\left(\hat{\rho}\right) 
=
1 - p_e
=
\frac{1}{2}
\left(
1 -
\left\langle {\sigma}_3 \right\rangle_\text{obs}
\right)
,
\ee
and the \textit{coherence}
\be\lb{e-coher}
\left\langle {\sigma}_+ \right\rangle_\text{obs}
\equiv
{\rm tr}\left(\hat{\rho}\hat{\sigma}_+ \right)/ {\rm tr}\left(\hat{\rho}\right) 
,
\ee
where 
$\rho_{i j}$
are the $i j$th components of the density matrix.
One can check that during the evolution the spin averages obey the
following identity  
\be
\left\langle {\sigma}_1 \right\rangle_\text{obs}^2
+
\left\langle {\sigma}_2 \right\rangle_\text{obs}^2
+
\left\langle {\sigma}_3 \right\rangle_\text{obs}^2
=
1 - 4 
\det (\hat\rho / \text{tr} \hat\rho)
%\frac{\det \hat\rho}{(\text{tr} \hat\rho)^2}
\leqslant 1
,
\ee
which means that for pure states the averages lie on the Bloch sphere
$
\left\langle {\sigma}_1 \right\rangle_\text{obs}^2
+
\left\langle {\sigma}_2 \right\rangle_\text{obs}^2
+
\left\langle {\sigma}_3 \right\rangle_\text{obs}^2
=
1$.

The unperturbed Hamiltonian of a quantum-optical two-level system is usually a linear combination
of the operators $|g\rangle \langle g|$ and $|e\rangle\langle e|$.
Up to an additive constant it equals to
\be\lb{eq:ham0}
\hat H_0 = \frac{1}{2} \hbar \omega_0 \hat{\sigma}_3
,
\ee
where $ \omega_0 $ is the transition frequency.
If the system is put into contact with a monochromatic electromagnetic wave of frequency
$ \omega_0 $ then in the leading order we can restrict ourselves with the dipole
interaction.
In the rotation-wave approximation (RWA) the corresponding Hamiltonian can be
reduced to the form
\be\lb{e:rabifreq}
\hat H_L = \frac{1}{2} \hbar \Omega 
\left(
\text{e}^{-i \omega_0 t}
\hat{\sigma}_+
+
\text{e}^{i \omega_0 t}
\hat{\sigma}_-
\right)
,
\ee
where $\Omega$ is the Rabi frequency which measures the strength of the interaction of the system's
dipole moment with the electromagnetic field,
and
$\hat{\sigma}_\pm =
\frac{1}{2} \left(\hat{\sigma}_1 \pm i \hat{\sigma}_2 \right)$.

For the purposes of simplifying the evolution equations it is often very convenient to perform a transition
from the \schrod picture to the interaction one.
One starts with the unitary transformation of the density operator
\be
\hat\rho_{(I)}
=
\text{e}^{i \hat H_0 t/\hbar}
\hat\rho \,
\text{e}^{-i \hat H_0 t/\hbar}
,
\ee 
where $\hat H_0$ is chosen as in (\ref{eq:ham0}).
This implies the transition formulae
\ba
&&
{\rm tr} \hat\rho_{(I)} = {\rm tr} \hat\rho
,\\&&
{\rm tr} (\hat\rho_{(I)} \hat O^{(I)} ) = {\rm tr} ( \hat\rho \hat O )
,\\&&
\hat O^{(I)}
=
\text{e}^{i \hat H_0 t/\hbar}
\hat O \,
\text{e}^{-i \hat H_0 t/\hbar}
,
\ea
where $\hat O$ refers to an observable's operator, the 
%sub- or superscript 
label $(I)$ indicates the interaction
picture with respect to $\hat H_0$, and absence of the label denotes the \schrod picture presentation.
Using the expressions above and Pauli matrices' properties,
we can write down the following transformation chart between the \schrod and interaction pictures to be used in the evolution equations
of the type (\ref{eq:master1}) or (\ref{eq:dtrho}):
\ba
&&
\hat{\rho}
\mapsto \hat{\rho}_{(I)}
,\
%\\&&\hat{H}_0 
\left[\hat{H}_0, \hat{\rho}\right] 
\mapsto 0
,\\&&
\hat H_L \mapsto 
%\hat H_L^{(I)} = 
\frac{1}{2} \hbar \Omega 
\left(
\hat{\sigma}_+
+
\hat{\sigma}_-
\right)
= \frac{1}{2} \hbar \Omega \hat{\sigma}_1
,\\&&
%\ea and also one can use the transformation formula \be
\vec\sigma
\mapsto
%\vec\sigma^{(I)} =
\left(
\begin{array}{ccc} 
\cos{(\omega_0 t)}& \sin{(\omega_0 t)} & 0\\ 
-\sin{(\omega_0 t)}& \cos{(\omega_0 t)} & 0\\
0&0 &1
\end{array}
\right)
\vec\sigma
,\lb{e:intpic-vecsig}\\&&
%\hat{\sigma}_3 \mapsto \hat{\sigma}_3,\
\hat{\sigma}_\pm \mapsto 
%\hat{\sigma}_\pm^{(I)}  =
\text{e}^{\pm i \omega_0 t} \hat{\sigma}_\pm
,
\ea
where by $\vec\sigma$ and $\vec\sigma^{(I)}$ we denote a set of the three Pauli operators in
the \schrod and interaction picture, respectively.

\section{Fourier transform in open quantum systems}\lb{s-a-ft}

Let us consider the following setup: some observable, $F (t)$, evolves according to quantum evolution equations.
Suppose that in absence of background effects its value is trivial: $F (t < 0) = f_c = \text{const}$.
Then at a certain moment of time, say $t = 0$, one ``switches on'' the background effects, such that the total function
becomes the following:
\be
F (t) = f_c \theta (- t) + f (t) \theta (t)
,
\ee
where $\theta$ is the Heaviside step function and $f (t) = F (t \geqslant 0)$.
For practical purposes we will be interested in the following two scenarios:

\begin{itemize}
%[(a)]%for small alpha-characters within brackets.
\item[(a)] 
function $f (t)$ tends to a constant value $f_\infty$ at $t \to +\infty$.

In this case the Fourier transform of the global function can be written as
\be
F (\omega)
=
\frac{1}{\sqrt{2 \pi}}
\left( f_c 
%+ f (0) 
+ f_\infty
\right) \delta (\omega)
+
[F (\omega)]_\text{reg}
,
\ee
where $[F (\omega)]_\text{reg}$ is the regular part of the Fourier transform:
\be\lb{e:ft-mon}
[F (\omega)]_\text{reg} =
\frac{1}{\sqrt{2 \pi}}
\int\limits_{0}^\infty
\left( f (t) - f_\infty \right) \text{e}^{i \omega t} d t
,
\ee
which is going to be the most informative for our purposes.

\item[(b)]
function $f (t)$ oscillates with a period $T$.

In this case the regular part of the Fourier transform of the global function can be computed as
\be\lb{e:ft-osc}
[F (n)]_\text{reg} =
\frac{1}{T}
\int\limits_{0}^T
f (t)  \text{e}^{2 \pi i (n / T) t} d t
,
\ee
where $n$ is an integer positive number.
\end{itemize}

%\newpage

\newpage


\begin{thebibliography}{99}

\bibitem{attal}
S. Attal and A. Joye, in:
\textit{Open Quantum Systems I: The Hamiltonian Approach}, 
eds. C.-A. Pillet (Springer, Berlin, 2006).

\bibitem{ilya}
A. Sergi, I. Sinayskiy, and F. Petruccione,
%``Numerical and Analytical Approach to the Quantum Dynamics
%of Two Coupled Spins in Bosonic Baths'',
Phys. Rev. A {\bf 80}, 012108 (2009).

\bibitem{gks76}
V. Gorini, A.  Kossakowski and E. C. G. Sudarshan,
%Completely positive dynamical semigroups of N-level systems. 
J. Math. Phys. \textbf{17}, 821-825 (1976).

\bibitem{lin76}
G. Lindblad,
%On the generator of quantum dynamical semigroups.
Commun. Math. Phys. \textbf{48}, 119-130 (1976).

\bibitem{bau66}
R. Bausch, 
%Bewegungsgesetze nicht abgeschlossener Quantensysteme.
Z. Phys. \textbf{193}, 246-265 (1966).

\bibitem{haa73}
F. Haake, 
%Statistical treatment of open systems by generalized master equations. 
Springer Tracts Mod. Phys. \textbf{66}, 98-168 (1973).

\bibitem{carbook}
H. J. Carmichael, \textit{An Open Systems Approach to Quantum Optics}, 
Lecture Notes in Physics (Springer-Verlag, Berlin, 1993).

\bibitem{gzbook}
C. W. Gardiner and P. Zoller, 
\textit{Quantum Noise} (Springer-Verlag, Berlin, 2000).

\bibitem{bpbook}
H.-P. Breuer and F. Petruccione, \textit{The Theory
of Open Quantum Systems} (Oxford University Press, 2002).

\bibitem{fesh58}
H. Feshbach, Ann. Phys. {\bf 5}, 357-390 (1958);
{\it ibid.} {\bf 19}, 287-313 (1962).

\bibitem{wong67}
J. Wong,
J. Math. Phys. \textbf{8}, 2039-2042 (1967).

\bibitem{faisa}
F. H. M. Faisal and J. V. Moloney,
J. Phys. B: At. Mol. Opt. Phys. {\bf 14}, 3603-3620 (1981).

\bibitem{datto}
G. Dattoli, A. Torre, and R. Mignani, Phys. Rev. A {\bf 42}, 1467-1475 (1990).

\bibitem{heg93}
G. C. Hegerfeldt,
Phys. Rev. A \textbf{47}, 449-455 (1993).

\bibitem{ang95}
S. Baskoutas \textit{et al.},
%, A. Jannussis, R. Mignani, and V. Papatheou,
J. Phys. A: Math. Gen. {\bf 26},  L819-L824 (1993);
P. Angelopoulou \textit{et al.},
%S. Baskoutas, A. Jannussis, R. Mignani, and V. Papatheou,
Int. J. Mod. Phys. B {\bf 9},  2083-2104 (1995).

\bibitem{rotter}
I. Rotter, arXiv:0711.2926; J. Phys. A {\bf 42}, 153001 (2009).

\bibitem{gsz08}
H. B. Geyer, F. G. Scholtz and K. G. Zloshchastiev,
in:
Proceedings of $12^\text{th}$ International Conference on
Mathematical Methods in Electromagnetic Theory
(Odessa, 2008)  250-252.

\bibitem{grsc11}
E.-M. Graefe and R. Schubert,
Phys. Rev. A \textbf{83}, 060101 (2011);
J. Phys. A \textbf{45}, 244033 (2012).

\bibitem{sz12}
A. Sergi and K. G. Zloshchastiev,
%Non-Hermitian quantum dynamics of a two-level system and models of dissipative environments
Int. J. Mod. Phys. B \textbf{27}, 1350163 (2013) [arXiv:1207.4877].

\bibitem{bg12}
D. C. Brody and E.-M. Graefe,
Phys. Rev. Lett. \textbf{109}, 230405 (2012) [arXiv:1208.5297].

\bibitem{ghk10}
E.-M. Graefe, M. H\"oning, and H. J. Korsch,
%Classical limit of non-Hermitian quantum dynamicsء generalized canonical structure
J. Phys. A {\bf 43}, 075306 (2010).

\bibitem{thila}
A. Thilagam, J. Chem. Phys. {\bf 136}, 065104 (2011).

\bibitem{sergi-mat}
A. Sergi,
%``Matrix Algebras in Non-Hermitian Quantum Mechanics'',
Comm. Theor. Phys. {\bf 56}, 96-98 (2011).

\bibitem{grmela}
M. Grmela, Phys. Lett. A {\bf 102}, 355-358 (1984).


\bibitem{gisin}
N. Gisin, J. Phys. A {\bf 14}, 2259-2267 (1981).

\bibitem{gisin2}
N. Gisin,
Physica A \textbf{111}, 364-370 (1982).

\bibitem{gisin3}
N. Gisin,
J. Math. Phys. \textbf{24}, 1779-1782 (1983).

\bibitem{ks87}
H. J. Korsch and H. Steffen, 
J. Phys. A {\bf 20}, 3787-3803 (1987).

%\cite{Yasue:1978bx,brash91}
\bibitem{various1}
M. D. Kostin, 
J. Chem. Phys. \textbf{57}, 3589-3591 (1972).

\bibitem{various2}
M. D. Kostin, 
J. Stat. Phys. \textbf{12}, 145-151 (1975).

\bibitem{various3}
I.~Bialynicki-Birula and J.~Mycielski,
%Nonlinear Wave Mechanics.
Annals Phys.\  {\bf 100}, 62-93 (1976).

\bibitem{various4}
K. Yasue,
%``Quantum Mechanics Of Nonconservative Systems,''
Annals Phys.\  {\bf 114}, 479-496 (1978).
  %%CITATION = APNYA,114,479;%%
	
\bibitem{various5}
N. A. Lemos,
%Dissipative forces and the algebra of operators in stochastic quantum mechanics,
Phys. Lett. A \textbf{78}, 239-241 (1980).
%\cite{brash91}
%\bibitem{brash91}

\bibitem{various6}
J. D. Brasher,
%information
Int. J. Theor. Phys. \textbf{30}, 979-984 (1991).

\bibitem{various7}
D. Schuch,
%Nonunitary connection between explicitly time-dependent and nonlinear approaches for the description of dissipative quantum systems
Phys. Rev. A \textbf{55}, 935-940 (1997).

\bibitem{various8}
M. P. Davidson,
%Comments on the nonlinear Schrodinger equation,
Nuov. Cim. B \textbf{116}, 1291-1294 (2001).

\bibitem{various9}
J. L. Lopez,
%Nonlinear Ginzburg-Landau-type approach to quantum dissipation,
Phys. Rev. E. \textbf{69}, 026110 (2004).

\bibitem{various10} K. G. Zloshchastiev, Grav. Cosmol. \textbf{16}, 288-297 (2010).

%\bibitem{various11} K. G. Zloshchastiev, Acta Phys. Polon. B \textbf{42} (2011) 261.

\bibitem{az11}
A.~V.~Avdeenkov and K.~G.~Zloshchastiev,
  %``Quantum Bose liquids with logarithmic nonlinearity: Self-sustainability and
  %emergence of spatial extent,''
J.\ Phys.\ B: At. Mol. Opt. Phys. {\bf 44}, 195303 (2011) [arXiv:1108.0847].
%  [arXiv:1108.0847 [cond-mat.quant-gas]].
  %%CITATION = JPBAB,B44,195303;%%K.~G.~Zloshchastiev, Eur. Phys. J. B \textbf{85} (2012) 273.
% [arXiv:1204.4652].

\bibitem{zlo2012}
K.~G.~Zloshchastiev, 
Eur. Phys. J. B \textbf{85}, 273 (2012) [arXiv:1204.4652].

% various1,various2,various3,various4,various5,various6,various7,various9,various9,various10,various11,az11


\end{thebibliography}
\end{document}